\documentclass[aps,prd,twocolumn,showpacs]{revtex4}
\usepackage{amsmath,amssymb}
\usepackage{hyperref}
\hypersetup{
    colorlinks,%
    citecolor=blue,%
    filecolor=blue,%
    linkcolor=magenta,%
    urlcolor=blue
}
\usepackage[all]{hypcap}
\usepackage{color}
\definecolor{gelb}{rgb}{0.7,0.5,0.2}
\newcommand{\bg}{{\mathbf{g}}}

\newcommand{\CC}{{\cal C}}

\newcommand{\CE}{{\cal E}}

\newcommand{\CK}{{\cal K}}
\newcommand{\CR}{{\cal R}}

\newcommand{\RZA}{\textrm{\tiny{RZA}}}

\def\a#1#2{\eta^{#1}_{\:\:#2}}
\def\ba#1{\boldsymbol{\eta}^{#1}}
\def\e#1#2{e_{#1}^{\:\:#2}}
\def\be#1{\mathbf{e}_{#1}}
\def\adot#1#2{\dot{\eta}^{#1}_{\:\:#2}}
\def\addot#1#2{\ddot{\eta}^{#1}_{\:\:#2}}

\def\one{\mathcal{O}(P)}

\def\t#1#2#3{#1^{\if#2- \else #2 \fi}_{\if#2- \else \:\: \fi #3}}
\def\tinv#1#2#3{#1^{\if#2- \else \:\: #2 \fi}_{\if#2- \else \fi #3}}
\def\tdot#1#2#3{\dot{#1}^{\if#2- \else #2 \fi}_{\if#2- \else \:\: \fi #3}}
\def\tddot#1#2#3{\ddot{#1}^{\if#2- \else #2 \fi}_{\if#2- \else \:\: \fi #3}}
\def\ttilde#1#2#3{\tilde{#1}^{\if#2- \else #2 \fi}_{\if#2- \else \:\: \fi #3}}

\begin{document}
\title{Lagrangian theory of structure formation in relativistic cosmology I:\\
Lagrangian framework and definition of a nonperturbative approximation}
\author{Thomas Buchert$^1$ and Matthias Ostermann$^{2,3}$}

\affiliation{$^1$Universit\'e de Lyon, Observatoire de Lyon, 
Centre de Recherche Astrophysique de Lyon, CNRS UMR 5574: Universit\'e Lyon~1 and \'Ecole Normale Sup\'erieure de Lyon, \\
9 avenue Charles Andr\'e, F--69230 Saint--Genis--Laval, France}

\affiliation{$^2$Arnold Sommerfeld Center, Ludwig--Maximilians--Universit\"at, Theresienstrasse 37, D--80333 M\"unchen, Germany}

\affiliation{$^3$Oskar--Maria--Graf Gymnasium,
Keltenweg 5, D--85375 Neufahrn, Germany\\
Emails: buchert@obs.univ-lyon.fr and mail@matthias-ostermann.de}

%\date{submitted ????? ??, 2003, accepted ????? ??, 200?}
%
% 98.80.-k Cosmology (see also 04 General relativity and gravitation;
% for origin and evolution of galaxies, see 98.62.A; for elementary
% particle and nuclear processes, see 95.30.C; for dark matter, see
% 98.80.Es Observational cosmology (including Hubble constant, distance scale, cosmological constant, early Universe, etc) 
% 98.80.Jk Mathematical and relativistic aspects of cosmology
% 04.20.-q Classical General Relativity
% 04.20.Cv Fundamental problems and general formalism
% 04.20.Dw Singularities and cosmic censorship 
% 04.25.Nx Post-Newtonian approximation; perturbation theory; related approximations 
%
%
%----------------------------------------------------------------%
%----------------------------------------------------------------%
\begin{abstract}
In this first paper we present a Lagrangian framework for the description of structure formation in general relativity, 
restricting attention to irrotational dust matter. As an application we present a self--contained derivation of a 
general--relativistic analogue of Zel'dovich's approximation for the description of structure formation in cosmology, 
and compare it with previous suggestions in the literature. This approximation is then investigated: paraphrasing the 
derivation in the Newtonian framework we provide general--relativistic analogues of the basic system of equations 
for a single dynamical field variable and recall the first--order perturbation solution of these equations. We then 
define a general--relativistic analogue of Zel'dovich's approximation and investigate its implications by functionally 
evaluating relevant variables, and we address the singularity problem. We so obtain a possibly powerful model that, 
although constructed through extrapolation of a perturbative solution, can be used to put into practice nonperturbatively, 
e.g. problems of structure formation, backreaction problems, nonlinear properties of gravitational radiation, and
light--propagation in realistic inhomogeneous universe models. With this model we also provide the key--building blocks for initializing 
a fully relativistic numerical simulation.
\end{abstract}

\maketitle

%----------------------------------------------------------------%
%----------------------------------------------------------------%
\section{Introduction}

General--relativistic analogues of the celebrated ``Zel'dovich approximation''
\cite{zeldovich:fragmentation,zeldovich:fragmentation2,zeldovich:rev,shandarin&zeldovich,sahni:approximation}
for the description of structure formation in the mildly nonlinear regime
have been suggested previously, first by Kasai in 1995 (\cite{kasai}; 
for extensions to second--order perturbation solutions see \cite{russ:rza},
\cite{russ:age}). We shall put these works into perspective, as well as those
by Matarrese and coworkers (\cite{matarrese&terranova,matarrese1}; 
for first-- and higher--order perturbation solutions see \cite{matarrese2}, 
\cite{matarrese3}), who discussed the relativistic analogues of the Newtonian equations 
in Lagrangian form. Croudace {\it et al.} and Salopek {\it et al.} discussed the Zel'dovich approximation in relation to 
spatial gradient expansion \cite{croudace,salopek}, and Ellis \& Tsagas \cite{ellis&tsagas},
proposed a covariant form for the peculiar--motion corresponding to Zel'dovich's ansatz.
The reader may also consult the seminal papers \cite{ellis:frames,maccallum} and \cite{henkuggla} 
that address the application of the orthonormal frame approach to relativistic cosmology. 

In this series of papers we reinforce the Lagrangian point of view with full 
rigor within the framework of Einstein's equations, keeping the 
formalism as close as possible to the Newtonian framework. 
In the present work we so obtain
(i) a natural analogue (in form and in spirit) of Zel'dovich's model
generalizing the approximation suggested by Kasai ({\it loc.cit.}), e.g.,
we obtain a quadratic form for the metric, useful for a realistic study of the light cone structure,
together with nontrivial projected curvature and Weyl tensor approximations, 
including a nonlinear gravitational radiation part;
(ii) general--relativistic Lagrangian equations that feature the 
Lagrange--Newton system of equations as a clear--cut geometrical limit 
(providing an alternative to the set of equations derived by Matarrese \& Terranova 
({\it loc.cit.}) by insisting on a single dynamical field variable; and (iii) a covariant description of relevant kinematical and dynamical
variables in the spirit of Ellis \& Tsagas ({\it loc.cit.}).
We shall also provide a number of useful details related to the basic equations and in particular to the 
electric and magnetic parts of the projected Weyl tensor that will be needed 
in forthcoming work.

Paraphrasing a previous Newtonian investigation \cite{buchert89,buchert92,buchert93,buchert:lagrangian,ehlersbuchert},
we look for general--relativistic analogues of (i) the Lagrangian deformation 
gradient of fluid elements, (ii) equations that feature this variable as the only 
dynamical one, and (iii) the corresponding first--order solution for perturbations at a FLRW background cosmology, which is then
extrapolated into the mildly nonlinear regime according to a definition that we shall 
provide here. We shall restrict our investigations to the matter model
``irrotational dust''. 
We obtain clear--cut answers to all of the above--mentioned points and
discover a very close analogy between relativistic and Newtonian equations
and models. This allows us to easily transfer ``Newtonian knowledge'' to the 
relativistic stage. The success of the corresponding Newtonian approximation 
suggests that the relativistic version of  Zel'dovich's model presented here is
a promising and possibly powerful one. 

This paper also aims at furnishing the basis for studies of
nonlinear perturbative and nonperturbative generic models that complement
standard perturbative studies and studies of exact solutions with high symmetry. 
This provides not only the basis for applications to structure formation in relativistic 
cosmology; future work will combine this approximation with exact evolution equations for 
the spatially averaged variables, \cite{buchert:onaverage1,buchert:onaverage2}, 
yielding nonperturbative models that are capable of addressing, e.g., the ``backreaction problem'' in 
relativistic cosmology, also as a possible source of  ``Dark Energy'', see the reviews \cite{ellisFOCUS,kolb:review,clarkson:review,darkenergy:review,buchert:focus,buchertrasanen,rasanenFOCUS}
and references therein. 
Furthermore, these relativistic models open the door to
other applications like, e.g., the understanding of nonlinear features of gravitational radiation, as well as 
of light--propagation and distance measurements in realistic inhomogeneous universe models that all cannot be addressed within the
Newtonian framework. Finally, since Newtonian simulations are often initialized with the Zel'dovich approximation, 
future fully relativistic simulations can be initialized by the relativistic form of this approximation as a first step.
The ingredients needed for realizing the initial conditions architecture of such simulations are provided here.

A major motivation of this line of works is to put ourselves into the position
to master a possible paradigm change in cosmology that entails the need for inhomogeneous
relativistic models. Curvature effects may play a key role in accessing 
the interpretation and high--precision determination of cosmological parameters in the near 
future. For example, the averaged spatial scalar curvature may evolve differently from a constant--curvature homogenous model, starting with
a small curvature as furnished by cosmic microwave background
observations and producing an effective negative curvature in the Late Universe \cite{rasanen:review,darkenergy:review,buchertcarfora,buchertrasanen},
bringing geometrical and topological  features into the fore. Another issue is the interpretation
of cosmological parameters that may be affected by curvature and Riemannian volume effects when 
comparing averaged variables in an inhomogeneous geometry with
averages on a Friedmannian template space (e.g., \cite{buchertcarfora:klingonletter,morphon:obs}). 
Finally, note that Newtonian cosmologies require periodic
boundary conditions for any model of structure formation \cite{buchertehlers}, which can be relaxed
in a relativistic setting (see e.g., \cite{buchert:jgrg,darkenergy:review,buchert:focus}).

\medskip

We proceed as follows.
In Section~\ref{sect:theoryN} we recall the Newtonian derivation of the Lagrangian equations together with Zel'dovich's approximation
in terms of a first--order Lagrangian perturbation solution. In Section~\ref{sect:theory} we paraphrase the Newtonian derivation within 
general relativity, give a compact analogous formulation
of Einstein's equations using Cartan's coframes, and discuss first--order perturbation solutions.
Section~\ref{sect:rza} defines the general--relativistic analogue of Zel'dovich's approximation 
and discusses it in full detail by functionally evaluating 
relevant variables including geometrical fields.
Section~\ref{sect:conclusions} proposes tests of the extrapolation into the nonlinear regime, discusses relations to the singularity problem, 
and highlights the main findings including follow--up prospects.
Appendixes are dedicated to alternative formulations of the governing equations and an example for the proposed approximation.

%----------------------------------------------------------------%
%----------------------------------------------------------------%
\section{Lagrangian theory of structure formation in Newtonian cosmology}
\label{sect:theoryN}

In this section we briefly recall the logical structure of a derivation
of Zel'dovich's model within Newtonian cosmology. Thereafter we shall contemplate
on Zel'dovich's original suggestion and his extrapolation idea in order to 
prepare ourselves for the relativistic setup.

\subsection{The Lagrange--Newton--System}

In the framework of Newtonian gravitation the field and evolution equations governing the motion of self--gravitating dust form a closed system in the Eulerian picture, consisting of the Eulerian evolution equations
\begin{eqnarray}
   \label{Euler1} \partial_t \vec{v} & = & - \left( \vec{v} \cdot \nabla \right) \vec{v} + \vec{g} \, , \\
   \label{Euler2} \partial_t \varrho & = & - \nabla \cdot \left( \varrho \vec{v} \right) ,
\end{eqnarray}
and the linear gravitational field equations
\begin{eqnarray}
   \label{Newton1} \nabla \times \vec{g} & = & \vec{0} \, , \\
   \label{Newton2} \nabla \cdot \vec{g} & = & \Lambda - 4 \pi G \varrho \; .
\end{eqnarray}
We call this system of equations the \emph{Euler--Newton--System} (ENS). Here, as usual, $\varrho$ is the dust's density, $G$ the gravitational and $\Lambda$ the cosmological constant.

Now we perform the transition from the Eulerian to the Lagrangian picture. Then, as we shall see, the trajectory field $\vec{x} = \vec{f}(\vec{X},t)$ -- defining the coordinate transformation at a fixed time, or a time--dependent diffeomorphism -- will be the only dynamical field variable that remains in the transformed equations, where $X^i$ are the Lagrangian coordinates, comoving with the fluid,
that are defined as to coincide with the Eulerian ones at some initial instant of time. The field $\vec{f}(\vec{X},t)$ measures the deviation of a fluid element's position at some time $t$ from its initial one, and its Lagrangian gradient $(f^i_{\;\,\vert j})$ measures the \emph{volume deformation} of fluid elements, where a vertical slash is used to denote partial derivative
with respect to Lagrangian coordinates. Upon introducing the trajectory field $\vec{f}(\vec{X},t)$ we implicitly solve the Eulerian evolution equations by
\begin{equation}
\vec v = \dot{\vec f} \;\;;\;\; \vec g = \ddot {\vec f} \;\;;\;\; \varrho = \frac{\mathring{\varrho}}{J}\;\;,\;\;J > 0\;\;,
\end{equation}
with the initial density field $\mathring{\varrho}(\vec  X)$, and $J$ the Jacobian determinant of the transformation from Eulerian to Lagrangian coordinates (using the functional 
determinant notation in the first expression),
\begin{equation}
   J \equiv \frac{\partial(f^1, f^2, f^3)}{\partial(X^1, X^2, X^3)} = \frac{1}{6} \epsilon_{ijk} \epsilon^{lmn} f^i_{\;\; \mid l} f^j_{\;\; \mid m} f^k_{\;\; \mid n} \; .
\end{equation}
The Eulerian field equations then assume the form of Lagrangian evolution equations, if the field strength is expressed through the trajectory field as above.
The resulting system of equations, the  \emph{Lagrange--Newton--System} (LNS), 
takes the following form (\cite{buchertgoetz} for $\Lambda = 0$ and \cite{buchert89} for $\Lambda \ne 0$):
\begin{eqnarray}
   \mathcal{J}(\ddot{f}^i , f^i , f^k) & = & 0 \; , \label{LNS1} \\
   \mathcal{J}(\ddot{f}^1 , f^2 , f^3) + \textrm{cycl.} & = & \Lambda J - 4 \pi G \mathring{\varrho} \; , \label{LNS2}
\end{eqnarray}
where $\mathcal{J}$ denotes the functional determinant of the expressions in brackets. 
Other forms of the Lagrange--Newton--System may be found in the review \cite{ehlersbuchert}, and in Appendix A.

Finally, we introduce the {\em Newtonian tidal tensor} $\CE_{ij}$,  
\begin{eqnarray}
   \CE^i_{\;\; j} \equiv g^i_{\;\, , j} - \frac{1}{3}\delta^i_{\;j}g^k_{\; , k} \qquad\qquad\nonumber\\ 
=\frac{1}{2J} \epsilon_{jkl} \mathcal{J}(\ddot{f}^i , f^k , f^l) + \frac{1}{3} \big( 4 \pi G \frac{\mathring{\varrho}}{J} - \Lambda \big) \delta^i_{\;j} \; ,
\end{eqnarray}
where we have inserted the field equation (\ref{Newton2}) in the second line. In terms of this form of the tidal tensor, written as a functional of $\vec f$,
we can express the LNS through the symmetry conditions on the tidal tensor $\CE_{ij}$:
\begin{equation}
\label{tidalformulationLNS}
   \CE_{[ij]} = 0 \qquad \textrm{and} \qquad \CE^k_{\;\; k} = 0 \; ,
\end{equation}
furnishing the four Lagrangian evolution equations for the three components of the trajectory field.

%----------------------------------------------------------------%
\subsection{Derivation of a first-order scheme and
Zel'dovich's approximation}

Now we proceed by linearizing Equations (\ref{LNS1}) and (\ref{LNS2}) at a reference background with respect to the deviations from this background: we assume the only variable $\vec{f}$ to be a superposition of a homogeneous and isotropic background deformation $\vec{f}_H(\vec{X},t) = a(t) \vec{X}$ and an inhomogeneous deformation field $\vec{p}(\vec{X},t)$, i.e.
\begin{equation}
   \vec{f}(\vec{X},t) = a(t) \vec{X} + \vec{p}(\vec{X},t) \; ,
\end{equation}
where for convenience $a(t_0) := 1$ and $\vec{p}(\vec{X},t_0) = 0$. It is sometimes useful to introduce the scaled trajectory field $\vec q = \vec F (\vec X ,t)\equiv  \vec{f}(\vec{X},t)/a(t)$ to describe motions
in a coordinate frame $\vec q$ that is comoving with the background solution. Correspondingly, we may also introduce the scaled deviation field 
$\vec{P}(\vec{X},t) \equiv \vec{p}(\vec{X},t)/a(t)$ that will be useful when comparing with the relativistic setting.

A homogeneous--isotropic deformation separately solves the LNS. This yields Friedmann's expansion law as a first integral, with background density $\varrho_H = \mathring{\varrho}_H a^{-3}$,
\begin{equation}
   H^2 \equiv \frac{\dot{a}^2}{a^2} = \frac{8 \pi G \varrho_H + \Lambda}{3} - \frac{\emph{const}}{a^2} \, .
\end{equation}

The first--order system of equations, which has to be solved, is
\begin{eqnarray}
  a^2 \nabla_0 \times \ddot{\vec{p}}  - \ddot{a} a \nabla_0 \times \vec{p} &\!\!=\!\!& \vec{0} \, , \label{firstorderN1} \\
  a^2 \nabla_0 \cdot \ddot{\vec{p}}+\left( 2 \ddot{a} a - a^2 \Lambda \right) \nabla_0 \cdot \vec{p} &\!\!=\!\!& -4 \pi G \left( \mathring{\varrho} - \mathring{\varrho}_H \right) . \;\;\; \label{firstorderN2}
\end{eqnarray}
Here $\nabla_0$ denotes derivative with respect to the Lagrangian coordinates. If we insert the field equations (\ref{Newton1}) and (\ref{Newton2}) at  initial time, we are able to express the source term in (\ref{firstorderN2}) by the divergence of the initial field--strength perturbation $\ddot{\vec{p}}(t_0)$. Now, we split the perturbation field $\vec{p}$ into a longitudinal part $\vec{p}^{\, L}$ and a transverse part $\vec{p}^{\, T}$. The resulting equations are for the transverse (divergence--free) part:
\begin{equation}
   \ddot{\vec{p}}^{\, T}-\frac{\ddot{a}}{a} \vec{p}^{\, T}  = \vec{0} \, , \label{firstorderN3}
\end{equation}
whereas the longitudinal (curl--free) part obeys:
\begin{equation}
   \ddot{\vec{p}}^{\, L} + \left( 2 \frac{\ddot{a}}{a} - \Lambda \right) \vec{p}^{\, L} = \frac{1}{a^2} \ddot{\vec{p}}^{\, L}(t_0)  \, . \label{firstorderN4} 
\end{equation}
Using the deviation field $\vec{P}(\vec{X},t)$,  Equations (\ref{firstorderN3}) and (\ref{firstorderN4}) take the form
\begin{eqnarray}
   \ddot{\vec{P}}^{\, T} + 2 H \dot{\vec{P}}^{\, T} &\!=\!& \vec{0} \, , \\
   \ddot{\vec{P}}^{L} + 2 H \dot{\vec{P}}^{L} - 4\pi G \varrho_H \vec{P}^{L} &\!=\!& \frac{1}{a^3} \vec{W}(\vec{X}) \: ,
\end{eqnarray}
where $\vec{W}(\vec{X}) \equiv \ddot{\vec{P}}^{L}(\vec{X},t_0) + 2 H(t_0) \dot{\vec{P}}^{L}(\vec{X},t_0)$ is the initial peculiar--acceleration field.
A more detailed derivation together with a general solution to these equations can be found in \cite{buchert92}. 
We wish to refer the reader also to \cite{ehlersbuchert} where the general equation and solution schemes for perturbations at any order can be found.

As a special case of the general first--order solution we obtain the ``Zel'dovich--approximation'', when we restrict the peculiar--velocity and peculiar--acceleration fields at some initial time $t_0$ by the ``slaving condition'':
\begin{equation}
\label{parallelity}
   \vec{u}(\vec{X},t) = \vec{w}(\vec{X},t) \, t \:\;;\;\; t = t_0 \;\;,
\end{equation}
where
\begin{equation}
   \vec{u} = a \dot{\vec{P}} \: , \quad \textrm{and} \quad \vec{w} = \dot{\vec u} + H \vec u = 2 \dot{a} \dot{\vec{P}} + a \ddot{\vec{P}}\;\;.
\end{equation}
The condition (\ref{parallelity}) is then preserved in time.

In the case of a spatially flat background without a cosmological constant the ``Zel'dovich--approximation'' reads
\begin{equation}
\label{NZA}
   \mbox{}^{\textrm{NZA}}\vec{F}(\vec{X},t) = \vec{X} + \frac{3}{2} \left[ \left( \frac{t}{t_0} \right)^{\frac{2}{3}} - 1 \right] \dot{\vec{P}}^{L}(\vec{X},t_0) \, t_0 \: .
\end{equation}
For general backgrounds including a constant--curvature term and a cosmological constant, see \cite{bildhaueretal}.

As we shall discuss in more detail later, Zel'dovich suggested to extrapolate this trajectory field into the mildly nonlinear regime, so that 
the nonlinearly evolved density can be calculated through its exact integral
\begin{equation}
\label{NZArho}
   \mbox{}^{\textrm{NZA}}\varrho = \frac{\varrho_H(t)}{\varrho_H(t_0)} \mathring{\varrho}(\vec{X}) \,/\, \mbox{}^{\textrm{NZA}}\! J_F (\vec{X},t) \, ,
\end{equation}
where $J_F \equiv \det (F^i_{\;\vert j})$ is evaluated for the comoving trajectory field (\ref{NZA}).
His motivation was that this expression for the density, if linearized at the background, coincides with the linearized solution for the density in comoving Eulerian coordinates $\vec q$, 
while the nonlinear expression is capable of describing a continuum that develops caustics in a finite time, in form similar to
the rectilinear motion of an inertial continuum \cite{zeldovichmyshkis} (for further discussions of this extrapolation idea and its subsequent developments see \cite{buchert89}).

We summarize the logical structure of the derivation of Zel'dovich's approximation: (i) the basic system of equations furnishes a closed system for a
single dynamical variable, the deformation field $\vec f$, or the deformation gradient $f^i_{\;\vert j}$; (ii) introducing a split into a background deformation (the Hubble flow) and a deviation field, we exploited the fact that we have only to linearize in the deformation field and not e.g. in the density deviations
as in the Eulerian picture; (iii) Eulerian fields, e.g. the density field, but also others, can then be evaluated as functionals of the linearized perturbation and
so provide nonlinear expressions as an extrapolation into the mildly nonlinear regime (i.e. up to shell--crossing singularities develop, after which the 
transformation of the Lagrangian functionals back to Eulerian space is no longer regular). The further restriction of initial data is not mandatory so that,
in principle, we can use this extrapolation idea also for the general first--order solution including vorticity. The functional for the vorticity is given by
Cauchy's exact integral \cite{buchert92}:
\begin{equation}
\vec\omega = \frac{{\vec\Omega} \cdot \nabla_0 {\vec F}}{a^2 J_F} \;\;;\;\;{\vec\Omega}\equiv {\vec\omega}({\vec X},t_0)\;.
\end{equation} 
%----------------------------------------------------------------%
\subsection{The strategy to find the corresponding relativistic approximation}

According to what has been said above, a general--relativistic analogue of Zel'dovich's approximation has to aim at (i) writing Einstein's equations in terms
of a system of evolution equations that all feature a single dynamical field variable corresponding to the Lagrangian deformation 
gradient; (ii) reducing constraint equations to constraints on initial data where possible; (iii) finding the general first--order solution of the system of evolution equations for the deformation variable, and then (iv) employing Zel'dovich's extrapolation idea to functionally express other variables in terms of the single perturbed deformation. It is clear that such a strategy results in a nonperturbative approximation of relevant field variables. For example, the resulting spatial metric
as a quadratic form of the deformation field will remain a quadratic form in this approximation. We are so able to keep highly nonlinear information encoded in the functional dependence on the perturbation variable (e.g. the exact density integral, the Ricci and Weyl curvatures etc.), while their 
solution is explicitly expressible in terms of constraint initial data and known time--dependent coefficients.  While Zel'dovich and his coworkers mainly exploited the
nonlinear functional dependence on the deformation in the density field, we here wish to apply this logic to all functionals of interest. 
As emphasized previously, this strategy is only applicable if the governing equations form a closed system for the deformation variable alone. 
%----------------------------------------------------------------%
%----------------------------------------------------------------%
\section{Lagrangian theory of structure formation in relativistic cosmology}
\label{sect:theory}

In this section we shall introduce the coframe field being the generalization of the Lagrangian deformation gradient  of Newtonian cosmology. In the general--relativistic case the deformation of fluid elements is no longer integrable, i.e.~instead of the basis $dx^a=\t{f}{a}{\mid i}dX^i$ we have to consider a non--exact basis $\ba{a}=\a{a}{i} dX^i$. While the linearly transformed (Lagrangian or local) basis in the Newtonian case derives from three functions (the components of the
trajectory field), here the linearly transformed local basis (here viewed in the cotangent space at a point of the manifold) involves nine functions (the coefficients of the set of coframe fields); hence we have to find at least
nine evolution equations.  
(We use latin letters $a,b, \dots = 1,2,3$ as counters in order to distingish them from coordinate indices $i,j, \dots = 1,2,3$; throughout the paper $\parallel$ denotes covariant derivative with respect to the 3--metric with a symmetric connection, whereas $\mid$ denotes partial derivative as before.) As in the previous section the three spatial deformation one--forms will be the only dynamical variables in our setup.

%----------------------------------------------------------------%
\subsection{The Lagrange--Einstein--System}

Restricting the matter model to ``irrotational dust'', the simplest spacetime foliation is given by a family of  flow--orthogonal hypersurfaces
with induced $3-$metric coefficients $g_{ij}$ in
the comoving and synchronous metric form,
\begin{equation}
^{(4)}\bg = - {\bf d}t \otimes {\bf d}t + \, ^{(3)} \bg \;\;;\;\; ^{(3)} \bg \equiv g_{ij}\, {\bf d}X^i \otimes {\bf d}X^j \, ,
\end{equation}
where $X^i$ are Gaussian normal (Lagrangian) coordinates that are constant
along flow lines (here geodesics). (The proper time--derivative is equal to the coordinate time--derivative, $u^{\mu}\frac{\partial}{\partial x^{\mu}}=\frac{\partial}{\partial t}$, below written with an overdot; Greek indices label spacetime and Latin ones spatial coordinates.) In our foliation we define the coefficients of the extrinsic curvature as usual,
\begin{equation}
   \CK_{ij} \equiv - \frac{1}{2} \dot{g}_{ij} \: .
\end{equation}
In the following we replace this quantity by the expansion tensor coefficients $\t{\Theta}{-}{(ij)} \equiv - \t{\CK}{-}{ij}$ together with the symmetry condition $\t{\Theta}{-}{[ij]} = 0$ (required in our foliation). In this frame the Einstein equations take the well--known form which has been introduced by Arnowitt, Deser and Misner \cite{ADM}, consisting of the evolution equations
\begin{eqnarray}
   \dot{\varrho} + \Theta \varrho & \!\!=\!\! & 0 \\
   \tdot{\Theta}{i}{j} + \Theta \t{\Theta}{i}{j} & \!\!=\!\! & \left( 4 \pi G \varrho + \Lambda \right) \t{\delta}{i}{j} \!-\! \t{R}{i}{\!j} \: ,
\label{ADMorigthree}
\end{eqnarray}
where ${R}_{ij}$ are the Ricci tensor coefficients corresponding to the 3--metric, and $\varrho$ the dust density field, completed by the constraint equations
\begin{eqnarray}
   \t{\Theta}{-}{ij} - \t{\Theta}{-}{ji} & \!\!=\!\! & 0 \label{Thetasym}\\
   R + \Theta^2 - \t{\Theta}{k}{l} \t{\Theta}{l}{k} & \!\!=\!\! & 16 \pi G \varrho + 2 \Lambda \label{hamconstraint} \\
\t{\Theta}{k}{i \parallel k} - \t{\Theta}{-}{\parallel i} & \!\!=\!\! & 0 \: . \label{momentumconstraints}
\end{eqnarray}
The latter equations are the Hamilton constraint and the momentum constraints.

In the $(3+1)$--split the raising and lowering of indices does not commute with the time derivative, i.e. for any tensor $T_{ij}$ we find 
\begin{equation}
   \tdot{T}{-}{ij} = \left( \t{g}{-}{ik} \t{T}{k}{j} \right)\dot{\mbox{}} = \t{g}{-}{ik} \tdot{T}{k}{j} + 2 \t{\Theta}{-}{ik} \t{T}{k}{j} \, .
\end{equation}
In the same way the covariant derivative and time--derivative do not commute. To handle this we derive the following useful relation for the
symmetric connection coefficients (Christoffel symbols):
\begin{equation}
   \tdot{\Gamma}{i}{kl} = \t{\Theta}{i}{k \parallel l} + \t{\Theta}{i}{l \parallel k} - \t{g}{ij}{} \t{\Theta}{-}{kl \parallel j} \, .
\end{equation}
Using it we are able to rewrite the time--derivative of the 3--Ricci tensor coefficients through spatial derivatives of the expansion tensor coefficients,
\begin{equation}
   \tdot{R}{-}{ij} = \t{\Theta}{k}{i \parallel j \parallel k} + \t{\Theta}{k}{j \parallel i \parallel k} - \t{\Theta}{{\;\;\parallel k}}{\!\!\!ij \;\; \parallel k} - \t{\Theta}{-}{\parallel i \parallel j} \, . \label{riccitheta}
\end{equation}
We can thus recast the evolution equation for the expansion tensor (\ref{ADMorigthree}) into a form that only features expressions built from the expansion tensor and its derivatives (apart from the exactly integrable source $\varrho$):
\begin{eqnarray}
\label{ADMorigthreedot}
   \tddot{\Theta}{i}{j} & \!\!+\!\! & \tdot{\Theta}{-}{} \t{\Theta}{i}{j} + \t{\Theta}{-}{} \tdot{\Theta}{i}{j} + 2 \t{\Theta}{i}{k} \tdot{\Theta}{k}{j} + 2 \t{\Theta}{-}{} \t{\Theta}{i}{k} \t{\Theta}{k}{j}  \nonumber\\
   & \!\!=\!\! & \t{\Theta}{{i\;\parallel k}}{j\; \parallel k} + \t{\Theta}{\parallel i}{\parallel j} - \t{\Theta}{{k\;\parallel i}}{j\; \parallel k} - \t{\Theta}{{i\;\;\;\; \parallel k}}{k \parallel j} \nonumber \\
   & \!\!+\!\! & 2 \left( 4 \pi G \varrho + \Lambda \right) \t{\Theta}{i}{j} - 4 \pi G \varrho \Theta \t{\delta}{i}{j} \: .
\end{eqnarray}
We note already that the trace of the expansion tensor is $\Theta = \dot{J}/J$, with $J$ given below, Eq.~(\ref{jacob}), so that we can immediately solve the continuity equation $\dot{\varrho} + \Theta \varrho = 0$ by integration and get the general integral in analogy to the Newtonian case:
\begin{equation}
\label{generalrho}
   \varrho = \mathring{\varrho} \frac{\mathring{J}}{J} \: .
\end{equation}
It is also useful to note that, in view of the momentum constraints (\ref{momentumconstraints}), we have $\t{g}{kl}{} \tdot{R}{-}{kl}=0$ and therefore
\begin{equation}
   \tdot{R}{-}{} = - 2 \t{\Theta}{k}{l} \t{R}{l}{k} \, .
\end{equation}

%----------------------------------------------------------------%
\subsubsection{Using Cartan's coframe fields}

Introducing Cartan's coframes $\ba{a}$ which define, up to rotations, a noncoordinate basis of 3--dimensional space we rewrite the spatial part of the metric as
\begin{equation}
   ^{(3)} \bg = \t{\delta}{-}{ab} \ba{a} \otimes \ba{b} \quad \Longrightarrow \quad g_{ij} = \delta_{ab} \a{a}{i} \a{b}{j} \: . \label{g_eta}
\end{equation}
Noncoordinate indices are raised and lowered by $\t{\delta}{-}{ab}$. Our choice is to simplify calculations by putting all the information on the initial data into the coframes by $\a{a}{i}(t_0)= \mathring{\eta}^{a}_{\;\; i}$. However, one could choose the more general orthogonal (and not orthonormal) matrix $\t{G}{-}{ab}$ instead of $\t{\delta}{-}{ab}$; then the coframe would take a simple form at some initial time, $^G\a{a}{i}(t_0)=\t{\delta}{a}{i}$, which would
formally come closer to the Newtonian description. We shall keep the standard definition throughout this first paper, but we shall come back to the other choice in forthcoming papers. Chandrasekhar \cite{chandra} discusses circumstances in which such a more general choice is useful.

Throughout this paper we define the Levi--Civita--tensor density by $\t{\epsilon}{-}{i_1i_2i_3}=(-1)^P$, where $P$ is the sign of the permutation $(1,2,3)\rightarrow(i_1i_2i_3)$ and $\t{\epsilon}{-}{i_1i_2i_3}=0$ if any two indices are the same. 

The determinant of the transformation between the coordinate and noncoordinate basis is given by
\begin{equation}
\label{jacob}
   J = \frac{1}{6} \epsilon_{abc} \epsilon^{ikl} \a{a}{i} \a{b}{k} \a{c}{l} \: .
\end{equation}
The (inverse) orthonormal vector basis is described by the triads (frames) $\be{a} = \e{a}{i} \partial / \partial X^i$, which can be expressed in terms of the coframes as follows,
\begin{equation}
\label{frametrafo}
   \e{a}{i} \a{a}{j} = \t{\delta}{i}{j} \quad \Longrightarrow \quad \e{a}{i} = \frac{1}{2J} \epsilon_{abc} \epsilon^{ikl} \a{b}{k} \a{c}{l} \: .
\end{equation}
Thus, the coefficients of the inverse metric take the form
\begin{equation}
   g^{ij} = \delta^{ab} \e{a}{i} \e{b}{j}  = \frac{1}{2J^2} \delta_{ce} \delta_{df} \epsilon^{ikl} \epsilon^{jmn} \a{c}{k} \a{d}{l} \a{e}{m} \a{f}{n} \: . \label{g_inv_eta}
\end{equation}

We rewrite the vanishing of the covariant derivative of the metric $\t{g}{kl}{}\t{g}{-}{kl\parallel i} = 0$ using frames and coframes and get our first two constraints (for the second one we apply $\a{a}{[i\parallel j]} = \a{a}{[i\mid j]}$),
\begin{eqnarray}
   \e{a}{k} \a{a}{k\parallel i} &\!\!=\!\!& 0 \, , \label{eta_der} \\
   \e{a}{k} \a{a}{i\parallel k} &\!\!=\!\!& 2 \e{a}{k} \a{a}{[i\mid k]} \, .
\end{eqnarray}
Since $J=\sqrt{g}$ in our choice of coordinates, where $g$ is the determinant of the 3--metric, we have to keep in mind that $J$ is not a scalar but a scalar density, i.e.~its covariant derivative vanishes according to (\ref{eta_der}),
\begin{equation}
   \frac{J_{\parallel i}}{J} = \frac{1}{2J} \epsilon_{abc} \epsilon^{mkl} \a{a}{m \parallel i} \a{b}{k} \a{c}{l} = \e{a}{k} \a{a}{k\parallel i} = 0 \, ,\label{jderiv1}
\end{equation}
but its partial derivative with respect to the spatial coordinates does not,
\begin{equation}
   \frac{J_{\mid i}}{J} = \e{a}{k} \a{a}{k\mid i} = \t{\Gamma}{k}{ki} \neq 0 \: . \label{jderiv2}
\end{equation}
It is also important to note that, contrary to the Newtonian case, the initial transformation determinant does not equal to one, $\mathring{J}\neq 1$ (as it would do for $\t{\delta}{-}{ab} \rightarrow \t{G}{-}{ab}$). The invariant volume element then is described by the tensor
\begin{equation}
   \t{\varepsilon}{-}{ikl} = J \t{\epsilon}{-}{ikl} \quad \textrm{and} \quad \t{\varepsilon}{ikl}{} = \frac{1}{J} \t{\epsilon}{ikl}{} \: . \label{invvol}
\end{equation}
Since we use an orthonormal noncoordinate basis, we simply have $\t{\varepsilon}{-}{abc} = \t{\epsilon}{-}{abc}$. We now turn to rewriting the ADM equations in terms of the coframes only. The expansion tensor coefficients with one index inverted, $\t{\Theta}{i}{j}$, will provide the closest analogy to the Newtonian case:
\begin{equation}
   \t{\Theta}{i}{j} = \adot{a}{j} \e{a}{i} = \frac{1}{2J} \epsilon_{abc} \epsilon^{ikl} \adot{a}{j} \a{b}{k} \a{c}{l} \: ,
\end{equation}
and the symmetry condition (\ref{Thetasym}) becomes
\begin{equation}
   \t{\Theta}{-}{[ij]} = \t{\delta}{-}{ab} \adot{a}{[i} \a{a}{j]} = 0 \: .
\end{equation}
Now, the three latter constraint equations have become evolution equations.

We define the 3--Riemann curvature tensor via the commutation relation of second covariant spatial derivatives, 
$ \a{a}{i \parallel k \parallel l} - \a{a}{i \parallel l \parallel k} = R^{\;\;\;\;m}_{kli}\a{a}{m}$, so we get: 
\begin{equation*}
   R_{klij} = \delta_{ab} ( \a{a}{i \parallel k \parallel l} - \a{a}{i \parallel l \parallel k} ) \a{b}{j} \;,
\end{equation*}
and, finally,
\begin{eqnarray}
   R^{i}_{\;jkl} &\!\!=\!\!& \e{a}{i} ( \a{a}{j \parallel k \parallel l} - \a{a}{j \parallel l \parallel k} ) \nonumber \\
   &\!\!=\!\!& \frac{1}{2J} \epsilon_{abc} \epsilon^{imn} \left( \a{a}{j \parallel k \parallel l} - \a{a}{j \parallel l \parallel k} \right) \a{b}{m} \a{c}{n} \;. \label{Riemann}
\end{eqnarray}
Contraction yields the Ricci tensor, i.e., $R_{ij} = R^{k}_{\;\;\;ikj}$,
\begin{equation}
   \t{R}{-}{ij} = \frac{1}{2J} \epsilon_{abc} \epsilon^{kmn} \left( \a{a}{j \parallel k \parallel i} - \a{a}{j \parallel i \parallel k} \right) \a{b}{m} \a{c}{n} \: .
\end{equation}
Note that the by simplifying this expression using (\ref{frametrafo}) and the identity
\begin{eqnarray*}
\label{levicivitaidentity1}
\epsilon^{ijk}\epsilon_{abc} \;=\;\delta^i_{\;a}\delta^j_{\;b}\delta^k_{\;c}\;+\;
\delta^i_{\;b}\delta^j_{\;c}\delta^k_{\;a}\;+\;\delta^i_{\;c}\delta^j_{\;a}\delta^k_{\;b}\quad\nonumber\\
\qquad\qquad\;-\;
\delta^i_{\;b}\delta^j_{\;a}\delta^k_{\;c}\;-\;\delta^i_{\;c}\delta^j_{\;b}\delta^k_{\;a}\;-\;
\delta^i_{\;a}\delta^j_{\;c}\delta^k_{\;b}\;\;,\,
\end{eqnarray*}
we get the alternative expression
\begin{equation}
   R_{ij} = \delta_{ab} ( \a{a\;\; \parallel k}{k \;\;\; \parallel i} - \a{a\;\;\;\;\; \parallel k}{k \parallel i} ) \a{b}{j} \;. \label{ricci_eta}
\end{equation}
Finally, we find for the mixed 3--Ricci tensor coefficients, expressed solely with the help of coframes,
\begin{equation}
   R^{i}_{\;\;j} = \delta_{ab} ( \a{a\;\; \parallel k \parallel i}{k} - \a{a\;\; \parallel i \parallel k}{k} ) \a{b}{j} \;,\label{Ricci}
\end{equation}
or, alternatively, with the help of frames (that often simplifies calculations),
\begin{equation}
   R^{i}_{\;\;j} = \delta^{ab} \e{a}{i} ( \e{b\;\; \parallel k \parallel j}{k} - \e{b\;\; \parallel j \parallel k}{k} )\;. \label{Ricci2}
\end{equation}
Contracting the Ricci tensor, i.e., $R = R^{k}_{\;\;k}$, we obtain for the scalar curvature in terms of coframes,
\begin{equation}
   R = \delta_{ab} ( \a{a\;\; \parallel l \parallel k}{k} - \a{a\;\; \parallel k \parallel l}{k} ) \a{b}{l} \;,\label{R}
\end{equation}
and, with (\ref{Ricci2}), in terms of frames,
\begin{equation}
   R = \delta^{ab} \e{a}{k} ( \e{b\;\; \parallel l \parallel k}{l} - \e{b\;\; \parallel k \parallel l}{l} ) \;.\label{R2}
\end{equation}

Using the coframe as the single dynamical variable the ADM equations become:
\begin{equation}
   \frac{1}{2}  \left( \t{\epsilon}{-}{abc} \t{\epsilon}{ikl}{} \adot{a}{j} \a{b}{k} \a{c}{l} \right)\dot{} = \left( 4 \pi G\mathring{J} \mathring{\varrho} + \Lambda J \right) \t{\delta}{i}{j} - J \t{R}{i}{\!j} \:, \label{ADMthree}
\end{equation}
and the set of (former) constraint equations become
\begin{eqnarray}
   \t{\delta}{-}{ab} \addot{a}{[i} \a{b}{j]} & \!\!=\!\! & 0 \label{ADMone} \;;\\
   \t{\epsilon}{-}{abc} \t{\epsilon}{mkl}{} \adot{a}{m} \adot{b}{k} \a{c}{l} & \!\!=\!\! & 16 \pi G \mathring{J} \mathring{\varrho} + 2 \Lambda J - J R \;;\quad \label{ADMtwo} \\
\left( \t{\epsilon}{-}{abc} \t{\epsilon}{ikl}{} \adot{a}{j} \a{b}{k} \a{c}{l} \right)_{\!\parallel i} & \!\!=\!\! & \left( \t{\epsilon}{-}{abc} \t{\epsilon}{ikl}{} \adot{a}{i} \a{b}{k} \a{c}{l} \right)_{\!\parallel j} \: .\label{ADMfour}
\end{eqnarray}
The first of these equations, (\ref{ADMone}), arises as the time--derivative of the symmetry condition for the expansion tensor, (\ref{ADMtwo}) comes from the Hamilton constraint, and (\ref{ADMfour}) represent the momentum constraints. We still use the covariant derivative and the Ricci tensor in this (overdetermined) system of $13$ evolution equations for the $9$ components of the deformation coefficients $\eta^a_{\;\;i}$, but do that only for the sake of readability. It is possible to express these in terms of the coframes only, too, as done above for the Ricci tensor and the scalar curvature. (We only have to make sure that
all these equations including the covariant derivatives could be expressed in terms of the coframes only.)
In this sense, an interesting form of (\ref{ADMthree}) is the following:
\begin{eqnarray}
   && \addot{a}{i} + \frac{1}{J} \t{\epsilon}{-}{bcd} \t{\epsilon}{mkl}{} \adot{a}{[i} \adot{b}{m]} \a{c}{k} \a{d}{l} - \Big( 4 \pi G\frac{\mathring{J} \mathring{\rho}}{J} + \Lambda \Big) \a{a}{i} \nonumber \\
   && = \frac{1}{J} \t{\delta}{ab}{} \t{\epsilon}{-}{bcd} \t{\epsilon}{mkl}{} \Big( \a{c}{k \parallel i \parallel m} - \a{c}{k \parallel m \parallel i} \Big) \a{d}{l} \;.
\label{ADM3alternative}
\end{eqnarray}
Here the equation is solely expressed in terms of the coframes and their time and spatial (covariant) derivatives.

We now condense this system of equations in a more compact form. First, we get another convenient form of (\ref{ADMthree}) when we split it into its trace and trace--free parts. With the Hamilton constraint the trace takes the form 
\begin{equation}
   \frac{1}{2} \t{\epsilon}{-}{abc} \t{\epsilon}{ikl}{} \addot{a}{i} \a{b}{k} \a{c}{l} = \Lambda J - 4 \pi G \mathring{J} \mathring{\varrho} \: , \label{LEStrace}
\end{equation}
which is Raychaudhuri's equation. The trace--free part on the other side is
\begin{eqnarray}
    \frac{1}{2} \Big( \t{\epsilon}{-}{abc} \t{\epsilon}{ikl}{} \addot{a}{j} \a{b}{k} \a{c}{l} &\!\!-\!\!& \frac{1}{3} \t{\epsilon}{-}{abc} \t{\epsilon}{mkl}{} \addot{a}{m} \a{b}{k} \a{c}{l} \t{\delta}{i}{j} \Big) \nonumber \\
    + \Big( \t{\epsilon}{-}{abc} \t{\epsilon}{ikl}{} \adot{a}{j} \adot{b}{k} \a{c}{l} &\!\!-\!\!& \frac{1}{3} \t{\epsilon}{-}{abc} \t{\epsilon}{mkl}{} \adot{a}{m} \adot{b}{k} \a{c}{l} \t{\delta}{i}{j} \Big) \nonumber \\
    &\!=\!& - J \t{\tau}{i}{\!j} \: , \label{LEStracefree}
\end{eqnarray}
where ${\tau}_{ij}$ are the coefficients of the trace--free part of the Ricci tensor to be calculated from Eq.~(\ref{ricci_eta}) and Eq.~(\ref{R}). We can now recast the set of equations (\ref{ricci_eta}), (\ref{ricci_eta}), (\ref{ADMone}) -- (\ref{LEStracefree}) into what we call the \emph{Lagrange--Einstein System for dust} (LES).
%----------------------------------------------------------------%
\subsubsection{Summary: the Lagrange--Einstein--System}

The following (over--determined) system of $13$ evolution equations for the $9$ coframe coefficient functions is equivalent to the ADM set of equations for the matter model `irrotational dust' (recall that this latter restriction implies that Eq.~(\ref{LES_E1}) already holds for the first time--derivative of the coframes):
\begin{eqnarray}
 \t{\delta}{-}{ab} \addot{a}{[i} \a{b}{j]}  &\!=\!&  0 \label{LES_E1} \\
\frac{1}{2} \t{\epsilon}{-}{abc} \t{\epsilon}{ikl}{} \addot{a}{i} \a{b}{k} \a{c}{l} &\!=\!& \Lambda J - 4 \pi G \mathring{J} \mathring{\varrho} \label{LES_E2} \\
 \left( \t{\epsilon}{-}{abc} \t{\epsilon}{ikl}{} \adot{a}{j} \a{b}{k} \a{c}{l} \right)_{\!\parallel i} &\!=\!& \left( \t{\epsilon}{-}{abc} \t{\epsilon}{ikl}{} \adot{a}{i} \a{b}{k} \a{c}{l} \right)_{\!\parallel j} \label{LES_H1} \\
 \t{\epsilon}{-}{abc} \t{\epsilon}{mkl}{} \adot{a}{m} \adot{b}{k} \a{c}{l} &\!=\!& 16 \pi G \mathring{J} \mathring{\varrho} + 2 \Lambda J - J R \label{LES_H3} \\
  \frac{1}{2} \Big( \t{\epsilon}{-}{abc} \t{\epsilon}{ikl}{} \addot{a}{j} \a{b}{k} \a{c}{l} &\!\!-\!\!& \frac{1}{3} \t{\epsilon}{-}{abc} \t{\epsilon}{mkl}{} \addot{a}{m} \a{b}{k} \a{c}{l} \t{\delta}{i}{j} \Big) \nonumber \\
  + \Big( \t{\epsilon}{-}{abc} \t{\epsilon}{ikl}{} \adot{a}{j} \adot{b}{k} \a{c}{l} &\!\!-\!\!& \frac{1}{3} \t{\epsilon}{-}{abc} \t{\epsilon}{mkl}{} \adot{a}{m} \adot{b}{k} \a{c}{l} \t{\delta}{i}{j} \Big) \nonumber \\
  &\!=\!& - J \t{\tau}{i}{\!j}\;. \label{LES_H2}
\end{eqnarray}
This Lagrange--Einstein--system is a system of equations described solely in terms of the coframes. We did not explicitly insert the trace--free part and trace of the 3--Ricci curvature into the above equations because the resulting equations are tedious to read. In principle it can be done with the equations given in this section. Of course, the covariant derivative can also be expressed in the coframes language by calculating the Christoffel symbols with (\ref{g_eta}) and (\ref{g_inv_eta}).

%----------------------------------------------------------------%
\subsubsection{Formulation with the Weyl tensor}

We are now going to reexpress the above Lagrange--Einstein--System in terms of parts of the projected Weyl tensor in order to furnish the analogy with
the tidal formulation of the Lagrange--Newton--System (\ref{tidalformulationLNS}). Here, the electric part of the Weyl tensor plays the role of the tidal tensor of Newtonian theory, whereas its magnetic part carries additional information and describes gravitomagnetic effects.

The Weyl tensor is defined as the trace--free part of the 4--Riemann curvature tensor,
\begin{equation}
   \t{C}{{\mu\nu}}{\;\;\kappa\lambda} = \mbox{}^{(4)}\!\t{R}{{\mu\nu}}{\;\;\kappa\lambda} - 2 \t{\delta}{[\mu}{[\kappa} \mbox{}^{(4)}\!\t{R}{\nu]}{\lambda]} + \frac{1}{3} \t{\delta}{[\mu}{[\kappa} \t{\delta}{\nu]}{\lambda]} \mbox{}^{(4)}\!R \, .
\end{equation}
It has $10$ independent components and thus carries all the information of the system. The Weyl tensor satisfies all of the symmetry conditions of the 4--curvature tensor, and in addition is trace--free over any two indices. It can be irreducibly split into two parts, called the electric and magnetic parts. Both parts are symmetric, trace--free tensors and have five independent components each,
\begin{equation}
   \t{E}{-}{\mu\nu} = \t{C}{-}{\mu\kappa\nu\lambda} u^{\kappa} u^{\lambda} \quad \textrm{and} \quad \t{H}{-}{\mu\nu} = \frac{1}{2} \epsilon_{\varrho\tau\kappa(\mu}{C}^{\varrho\tau}_{\;\;\;\nu)\lambda} u^{\kappa} u^{\lambda} \, , \label{weylparts}
\end{equation}
After the $(3+1)$--split the electric (tidal) part $\t{E}{i}{j}$ of the Weyl tensor and its magnetic part $\t{H}{i}{j}$ take the following forms:
\begin{eqnarray}
   \t{E}{i}{j} &\!\!=\!\!& - \tdot{\Theta}{i}{j} - \t{\Theta}{i}{k} \t{\Theta}{k}{j} - \frac{1}{3} \Big( 4 \pi G \frac{\mathring{\varrho}\mathring{J}}{J} -  \Lambda \Big) \t{\delta}{i}{j} \, , \label{ele0}\\
   \t{H}{i}{j} &\!\!=\!\!& - \frac{1}{J} \t{\epsilon}{ikl}{} \t{\Theta}{-}{jk \parallel l} \label{mag0}\, .
\end{eqnarray}
We also note the useful expressions
\begin{eqnarray}
\label{Esigma}
\t{E}{i}{j} = - \tdot{\sigma}{i}{j} - \Big( \t{\Theta}{i}{k} \t{\Theta}{k}{j} - \frac{1}{3} \t{\Theta}{l}{k} \t{\Theta}{k}{l} \, \t{\delta}{i}{j} \Big)\nonumber \\
= - \tdot{\sigma}{i}{j} - \frac{2}{3} \Theta  \t{\sigma}{i}{j} - \Big( \t{\sigma}{i}{k} \t{\sigma}{k}{j}  - \frac{1}{3}  \t{\sigma}{l}{k} \t{\sigma}{k}{l}  \t{\delta}{i}{j} \Big)   \;,
\end{eqnarray}
where ${\sigma}_{ij}$ are the components of the shear tensor.

We rewrite the above parts of the Weyl tensor by fully expressing them through coframes,  
\begin{eqnarray}
   \t{E}{i}{j} &\!\!=\!\!& - \frac{1}{2J} \t{\epsilon}{-}{abc} \t{\epsilon}{ikl}{} \addot{a}{j} \a{b}{k} \a{c}{l} - \frac{1}{3} \Big( 4 \pi G \frac{\mathring{\varrho}\mathring{J}}{J} -  \Lambda \Big) \t{\delta}{i}{j}, \quad \label{ele}\\
   \t{H}{i}{j} &\!\!=\!\!& - \frac{1}{J} \t{\delta}{-}{ab} \t{\epsilon}{ikl}{} \big( \adot{a}{j\parallel l} \a{b}{k} + \adot{a}{j} \a{b}{k \parallel l} \big). \label{mag}
\end{eqnarray}
We infer that the projected electric part of the Weyl tensor yields a direct generalization of the tidal formulation of the Lagrange--Newton--System 
(\ref{tidalformulationLNS}):
\begin{eqnarray}
   \t{E}{-}{[ik]} = 0 & \Longleftrightarrow & (\ref{LES_E1}) \: , \label{Esym} \\
   \t{E}{k}{k} = 0 & \Longleftrightarrow & (\ref{LES_E2}) \: , \label{Etrace}
\end{eqnarray}
whereas the magnetic part reproduces the momentum constraints and again the symmetry condition for the time--derivative of the expansion tensor:
\begin{eqnarray}
   \t{H}{-}{[ik]} = 0 & \Longleftrightarrow & (\ref{LES_H1}) \: , \label{Hsym} \\
   \t{H}{k}{k} = 0 & \Longleftarrow & (\ref{LES_E1}) \: . \label{Htrace}
\end{eqnarray}
At this stage the symmetry conditions on the electric and magnetic parts of the Weyl tensor do not cover  all of the equations of the 
LES: the electric part of the Weyl tensor fully covers the `electric part' of the LES, Eqs.~(\ref{LES_E1}) and (\ref{LES_E2}) -- which we need for the translation of the Newtonian approximation --, while the magnetic
part of the Weyl tensor just captures part of the `magnetic part' of the LES, namely Eqs.~(\ref{LES_H1}), i.e. in total $7$ equations.
We believe that another form of the magnetic part could eventually provide a symmetric formulation of the whole system, but we did not succeed to find it.

%----------------------------------------------------------------%
\subsubsection{A geometrical Newtonian limit for spatial deformations}

It is easy to confirm that we obtain the LNS as the following geometrical (spatial) limit to the Lagrange--Einstein--System of equations. The transition from general 
coframe coefficients $\a{a}{i}$ to those of the integrable (Newtonian) form
\begin{equation}
   \a{a}{i} \quad \longrightarrow \quad ^N\a{a}{i} = \t{f}{a}{\mid i}
\end{equation}
directly transforms the equations (\ref{LES_E1}), (\ref{LES_E2}) into (\ref{LNS1}), (\ref{LNS2}). This is particularly easy to see using differential forms as done in Appendix A. We also find that, as expected, the electric part of the Weyl tensor reduces to (minus) the tidal tensor of the Newtonian picture,
$E_{ij} \longrightarrow - {\cal E}_{ij}$. The spatial line element then takes the well--known Newtonian form
\begin{equation}
   ^{(3)} \bg \;\rightarrow \;\,^N\! \t{g}{-}{ij} {\bf d}X^i \otimes {\bf d}X^j = \t{\delta}{-}{ab} \t{f}{a}{\mid i} \t{f}{b}{\mid j} {\bf d}X^i \otimes {\bf d}X^j \, ,
\end{equation}
i.e.~a Euclidean line element that was transformed using the transformation $\vec f$. The basis vectors (frames) orthonormal to the coframes in the Newtonian limit are $\tinv{h}{i}{,a} \equiv \mbox{}^{N\!} \e{a}{i}$ (where a comma denotes derivative with respect to Eulerian coordinates). They obey $\t{f}{a}{\mid j} \tinv{h}{i}{,a} = {\delta}_{j}^{\;i}$ and take the following form,
\begin{equation}
   \tinv{h}{i}{,a} = \frac{1}{2 \mbox{}^{N\!} J} \t{\epsilon}{-}{abc} \t{\epsilon}{ikl}{} \t{f}{b}{\mid k} \t{f}{c}{\mid l} \: .
\end{equation}

In this limit the connection coefficients reduce to the inertial force terms
\begin{equation}
   ^N \!\, \t{\Gamma}{i}{kl} = \t{f}{a}{\mid kl} \tinv{h}{i}{,a} = \frac{1}{2 \,^N\!J} \t{\epsilon}{-}{abc} \t{\epsilon}{imn}{} \t{f}{a}{\mid kl} \t{f}{b}{\mid m} \t{f}{c}{\mid n} \neq 0 \, .
\end{equation}
Although the Christoffel symbols do not vanish because in the Lagrangian picture of the Newtonian equations we adopt curvilinear coordinates, we can easily verify that $\mbox{}^{N\!} \t{R}{-}{ij} = 0$ as required and
\begin{equation}
   ^N \! \a{a}{k \parallel l} = \t{f}{a}{\mid kl} - \,^N \!\, \t{\Gamma}{i}{kl} \t{f}{a}{\mid i} = 0 \, .
\end{equation}
However, since covariant derivative and time--derivative do not commute, we have nonvanishing
\begin{equation}
   \mbox{}^{N\!} \adot{a}{k \parallel l} = \tdot{f}{a}{\mid kl} - \tdot{f}{a}{\mid i} \tinv{h}{i}{,b} \t{f}{b}{\mid kl} \: .
\end{equation}
Combining the latter equations we immediately confirm that the magnetic part of the Weyl tensor (see Eq.~(\ref{mag}) below) has no nontrivial Newtonian counterpart, as it always vanishes in the geometrical limit defined above,
\begin{equation}
   ^N \! \t{H}{i}{j} = - \frac{1}{\mbox{}^{N\!} J} \t{\delta}{-}{ab} \t{\epsilon}{imn}{} \t{f}{a}{\mid j} \big( \tdot{f}{b}{\mid mn} - \tdot{f}{b}{\mid k} \tinv{h}{k}{,c} \t{f}{c}{\mid mn} \big) = 0 \: .
\end{equation}
This geometrical limiting procedure is spatial and, therefore, does not involve a limit $c \rightarrow \infty$. The light cone structure is simply not seen within the $3-$space by comoving observers (the Lorentzian structure appears in time--direction only). Note that this limit explicitly demonstrates that a {\it Minkowskian limit} (consisting now of this geometrical limit {\it and}
$c \rightarrow \infty$) in the comoving--synchronous slicing of spacetime is well--defined. In a post--Newtonian formulation, the Minkowskian limit leads to
the Eulerian form of the Newtonian equations, while in this setting it leads to their Lagrangian form (see also \cite{ehlers,ehlersbuchert:weyl}).

%----------------------------------------------------------------%
\subsection{Derivation of a first--order scheme} \label{subsect:firstorder}

To derive a first--order perturbation scheme we choose a flat homogeneous and isotropic background with some initial perturbation thereof, $\t{P}{a}{i}(X,t_0) = \t{\mathring{P}}{a}{i}(X)$ (henceforth, we omit the vector symbol over $X$ for notational ease). With this choice of coordinates, the coframe coefficients take the form
\begin{equation}
   \a{a}{i} = a(t) \left[ \t{\delta}{a}{i} + \t{P}{a}{i}(X,t) \right] \, , \quad \textrm{and} \quad \t{\tilde{\eta}}{a}{i} \equiv \frac{1}{a} \a{a}{i} \: , \label{firstordereta}
\end{equation}
where $a(t)$ is the usual scale factor and $\t{P}{a}{i}(X,t)$ the inhomogeneous deviation (perturbation) field with respect to which we shall 
linearize the equations (for notational ease we shall {\em not} write $\mbox{}^{(1)}\t{P}{a}{i}$). We call $\t{\tilde{\eta}}{a}{i}$ the ``peculiar''--coframe. At some initial time we assume $a(t_0)=1$ and $\t{\mathring{P}}{a}{i}(X) \neq 0$. The initial perturbation cannot be set to zero because that would flatten the initial metric and suppress any metric evolution, as pointed out by Matarrese $\&$ Terranova \cite{matarrese&terranova} as well as 
Russ et al. \cite{russ:rza}.

Thus, the initial 3--Ricci tensor is not equal to zero. Remember that, generally, for a homogeneous and isotropic background within a space of constant intrinsic scalar curvature $-6 k/a^2$ the zeroth order of (\ref{ADMthree}) is $\mbox{}^{H}\! \t{R}{i}{j} = - 2 \frac{k}{a^2} \t{\delta}{i}{j}$, see for example \cite{kasai}, whereas we have $\mbox{}^{H}\! \t{R}{i}{j} = 0$. However, since we can choose appropriate initial perturbations ($\t{\mathring{R}}{i}{j} \neq 0$!) to describe the space we want, this choice implies no restriction of generality.

The perturbation $\t{P}{a}{i}$ only appears summed over the noncoordinate index in the equations, so we introduce the following tensor coefficients and their trace:
\begin{equation}
   \t{P}{i}{j} \equiv \tinv{\delta}{i}{a} \t{P}{a}{j} \quad \textrm{and} \quad P \equiv \t{P}{k}{k} = \tinv{\delta}{k}{a} \t{P}{a}{k} \;,
\end{equation}
and use this notation throughout the remaining part of this section to make reading more convenient. (Note that within the first--order scheme we have two true tensor indices here.)

%------------------------------------------------------------------------%
\subsubsection{Field variables and equations in the first--order scheme} 

Up to the first order the spatial metric takes the form
\begin{equation}
   \mbox{}^{(1)}\t{g}{-}{ij} = a^2 \left( \t{\delta}{-}{ij} + \t{P}{-}{ij} + \t{P}{-}{ji} \right) .
\end{equation}
(Recall that we here aim at {\it strict} linearization, not to be confused with the relativistic form of Zel'dovich's approximation that we shall define below). 

The determinant of the transformation from coordinates to the noncoordinate basis becomes
\begin{equation}
   \mbox{}^{(1)\!}J = a^3 \left( 1 + P \right) \quad \textrm{and} \quad \mbox{}^{(1)\!}\mathring{J} = \left( 1 + \mathring{P} \right) \neq 1 \, . \nonumber
\end{equation}

The first--order Christoffel symbols are
\begin{equation}
   \mbox{}^{(1)}\t{\Gamma}{i}{kl} = \left( \t{P}{i}{(k\mid l)} + \tinv{P}{{\;\;i}}{(k\;\;\mid l)} - \tinv{P}{{\;\;\;\;\mid i}}{(kl)} \right) , \label{gammafirstorder}
\end{equation}
and in particular, we find $\mbox{}^{(1)}\t{\Gamma}{k}{ki} = \t{P}{-}{\, \mid i}\,$.

To begin with the first--order LES equations, let us have a look at the symmetry condition (\ref{ADMone}). Straightforward calculation up to first order yields
\begin{equation}
   \tdot{P}{-}{[ij]} = 0 \quad \Longrightarrow \quad \t{P}{-}{[ij]} = \t{\mathring{P}}{-}{[ij]} \, . \label{sym1}
\end{equation}

We now derive the first--order expressions for the covariant derivative of the coframe coefficients and its time--derivative,
\begin{eqnarray}
   \mbox{}^{(1)}\tinv{\delta}{i}{a} \a{a}{k \parallel l} &\!=\!& a \left( \t{P}{i}{[k \mid l]} + \tinv{P}{{\;\;\;\,\mid i}}{(kl)} - \tinv{P}{{\;\,i}}{(k\;\; \mid l)} \right) , \nonumber \\
   \mbox{}^{(1)}\tinv{\delta}{i}{a} \adot{a}{k \parallel l} &\!=\!& a \tdot{P}{i}{k\mid l} + \dot{a} \left( \t{P}{i}{[k \mid l]} + \tinv{P}{{\;\;\;\,\mid i}}{(kl)} - \tinv{P}{{\;\,i}}{(k\;\; \mid l)} \right) . \nonumber
\end{eqnarray}
Thus, we find an expression for the momentum constraints (\ref{ADMfour}) up to first order, which are 
\begin{equation}
   \tdot{P}{k}{{\,[k \mid i]}} = 0 \quad \Longrightarrow \quad \t{P}{k}{{\,[k \mid i]}} = \t{\mathring{P}}{k}{{\,[k \mid i]}} \, . \label{momconstrfirstorder}
\end{equation}
Assuming the Hamilton constraint (\ref{ADMtwo}) holds for the homogeneous background, that is, for vanishing perturbations, we get
\begin{equation}
  3 H^2 = 8 \pi G \varrho_H + \Lambda \, , \label{hamconstrhom}
\end{equation}
where $\varrho_H = \varrho_{H0} / a^3$ is the homogeneous background density. 
Then the first-order Hamilton constraint reads
\begin{equation}
  H \dot{P} + 4 \pi G \varrho_H P = -\frac{1}{4} \mbox{}^{(1)\!} R \, . \label{hamconstrfirstorder}
\end{equation}
Next, we determine the Ricci tensor and its trace up to first order. Since the Christoffel symbols (\ref{gammafirstorder}) are of order $\one$, their product is always of second or higher order, and we find
\begin{eqnarray}
   \mbox{}^{(1)\!}\t{R}{-}{ij} &\!=\!& \mbox{}^{(1)}\t{\Gamma}{k}{ij\mid k} - \mbox{}^{(1)}\t{\Gamma}{k}{ki \mid j} \nonumber \\
   &\!=\!& \t{P}{-}{\mid ij} - \tinv{P}{\mid k}{ij\;\;\mid k} \label{RICfirstorder} \\
   &\!+\!& \big( \tinv{\mathring{P}}{{\;\;\; \mid k}}{(ik)\;\;\mid j} + \tinv{\mathring{P}}{{\;\;\;\,\mid k}}{(jk)\;\;\mid i} + \tinv{\mathring{P}}{{\;\;\; \mid k}}{[ij]\;\;\mid k} - 2 \t{\mathring{P}}{-}{\mid ij} \big) , \nonumber
\end{eqnarray}
where we used the two constraint equations (\ref{sym1}) and (\ref{momconstrfirstorder}) above to express most of the terms by means of the initial perturbation field.

Obviously, the first--order trace is solely dependent on the initial perturbation,
\begin{equation}
   \mbox{}^{(1)\!}R = - \frac{4}{a^2} \t{P}{k\;\;\mid l}{{\,[k\;\;\mid l]}} = - \frac{4}{a^2} \t{\mathring{P}}{k\;\;\mid l}{{\,[k\;\;\mid l]}} \: , \label{RSfirstorder}
\end{equation}
so the first-order Hamilton constraint finally yields
\begin{equation}
  H \dot{P} + 4 \pi G \varrho_H P = \frac{1}{a^2} \t{\mathring{P}}{k\;\;\mid l}{{\,[k\;\;\mid l]}} \: .
\end{equation}

Now, we address the six evolution equations of the LES (\ref{ADMthree}). The homogeneous equations reduce to the trace and, with (\ref{hamconstrhom}), yield Friedmann's acceleration law
\begin{equation}
  3 \frac{\ddot{a}}{a} = - 4 \pi G \varrho_H + \Lambda \, , \label{friedmann}
\end{equation}
where $\varrho_H = \varrho_{H0} / a^3$ again is the homogeneous background density. The first-order equations then are
\begin{equation}
   \tddot{P}{i}{j} + 3H \tdot{P}{i}{j} + H \dot{P} \t{\delta}{i}{j} + 4 \pi G \varrho_H P \t{\delta}{i}{j} = - \mbox{}^{(1)\!} \t{R}{i}{j} \, , \label{LESfirstorder}
\end{equation}
and using the Hamilton constraint (\ref{hamconstrfirstorder}) we get
\begin{equation}
   \tddot{P}{i}{j} + 3H \tdot{P}{i}{j} - \frac{1}{3} H \dot{P} \t{\delta}{i}{j} - \frac{4 \pi G}{3} \varrho_H P \t{\delta}{i}{j} = - \mbox{}^{(1)} \t{\tau}{i}{j} \, , \label{LESfirstorder_alt}
\end{equation}
where $\mbox{}^{(1)}\t{\tau}{i}{j} \equiv \mbox{}^{(1)}\!\t{R}{i}{j} - \frac{1}{3} \mbox{}^{(1)}\!R \t{\delta}{i}{j}$ represent the trace--free part of the 3--Ricci tensor. Hence, it is convenient to write these equations in their representation (\ref{LEStrace}) and (\ref{LEStracefree}), i.e.~split into trace and trace--free parts. The first--order Raychaudhuri equation takes the following form:
\begin{equation}
  \ddot{P} + 2 H \dot{P} - 4 \pi G \varrho_H P = 0 \, . \label{rayfirstorder}
\end{equation}

We get another very convenient form of the scalar equation up to first order, when we take the trace of (\ref{LESfirstorder}) and replace $3 H \dot{P}$ by applying the Hamilton constraint (i.e.~instead of replacing the Ricci scalar as done in (\ref{LESfirstorder_alt})). 

Since the first--order trace $\mbox{}^{(1)}\!R$ only depends on the initial perturbation field, we subtract the resulting equation at $t=t_0$ from the general one to get rid of it. Hence, the alternative form of the trace equation is 
\begin{equation}
   \ddot{P} + 3H \dot{P} = \frac{1}{a^2} \big( \ddot{P}(t_0) + 3 H(t_0) \dot{P}(t_0) \big) \, . \label{tracefirstorder2}
\end{equation}
Note that this equation is equivalent to (\ref{rayfirstorder}) (see the general solution in the next section). However, because of the following equations, this equation seems to be the more natural choice even if (\ref{rayfirstorder}) is well-known from Newtonian theory.

Now, the trace--free part of the set of evolution equations (\ref{LEStracefree}) is
\begin{equation}
   \tddot{\Pi}{i}{j} + 3H \tdot{\Pi}{i}{j} = - \mbox{}^{(1)} \t{\tau}{i}{j} \, ,
\end{equation}
where $\t{\Pi}{i}{j} \equiv \t{P}{i}{j} - \frac{1}{3} P \t{\delta}{i}{j}$.
This last set of equations governs gravitational radiation, see also \cite{matarrese&terranova}. This is made clear if we express $\mbox{}^{(1)}\t{\tau}{i}{j}$ by (\ref{RICfirstorder}) and (\ref{RSfirstorder}) and -- analogous to what we did with the alternative form of the trace equation -- choose the more convenient form
\begin{equation}
   \tddot{\Pi}{-}{ij} + 3H \tdot{\Pi}{-}{ij} + \frac{1}{a^2} \big( \t{\Pi}{-}{\mid ij} - \tinv{\Pi}{{\;\mid k}}{ij\;\;\mid k} \big) = \frac{1}{a^2} \t{\CC}{-}{ij} \, , \label{ADMthreefirstorder}
\end{equation}
where $\CC_{ij}$ are constants depending on the initial perturbations,   
\begin{equation}
   \t{\CC}{-}{ij} = \tddot{\Pi}{-}{ij}(t_0) + 3 H(t_0) \tdot{\Pi}{-}{ij}(t_0) + \t{\Pi}{-}{\mid ij}(t_0) - \tinv{\Pi}{{\; \mid k}}{ij\;\;\mid k}(t_0) \, . \nonumber
\end{equation}
We now determined six of the nine equations which govern the evolution of the perturbation in the first-order scheme. The remaining three equations are equations (\ref{sym1}), which also arise as antisymmetric part of (\ref{ADMthree}), if we understand these as a set of nine evolution equations.

%------------------------------------------------------------------------%
\subsubsection{Parts of the Weyl tensor in the first--order scheme}

First, we determine the electric part of the Weyl tensor up to the first order and find
\begin{equation}
   \mbox{}^{(1)}\t{E}{i}{j} = \mbox{}^{H\!}\t{E}{i}{j} - \tddot{P}{i}{j} - 2 H \tdot{P}{i}{j} + \frac{4\pi G}{3} \varrho_H P \t{\delta}{i}{j} \: ,
\end{equation}
where $\mbox{}^{H\!}\t{E}{i}{j}$ is the homogeneous part,
\begin{equation}
   \mbox{}^{H\!}\t{E}{i}{j} = - \frac{\ddot{a}}{a} \t{\delta}{i}{j} - \frac{1}{3} \left( 4\pi G \rho_H - \Lambda \right) \t{\delta}{i}{j} \: ,
\end{equation}
which reproduces Friedmann's acceleration law (\ref{friedmann}). As in the Newtonian analogue we are able to rewrite the first part of the ADM equations in terms of the electric part of the Weyl tensor, i.e.
\begin{eqnarray}
   ^{(1)}\t{E}{k}{k} = 0 & \Longleftrightarrow & (\ref{rayfirstorder}) \: , \\
   ^{(1)}\t{E}{-}{[ik]} = 0 & \Longleftrightarrow & (\ref{sym1}) \: .
\end{eqnarray}

On the other hand, using (\ref{sym1}) we find the magnetic part of the Weyl tensor to be
\begin{equation}
   \mbox{}^{(1)}\t{H}{i}{j} = - \frac{1}{a} \epsilon^{ikl} \tdot{P}{-}{jk \mid l} \: ,
\end{equation}
so $\t{H}{i}{j}$ has no homogeneous part, $\mbox{}^{H\!}\t{H}{i}{j}=0\,$. The vanishing of its trace reproduces (\ref{sym1}), whereas the vanishing of the antisymmetric part reproduces the momentum constraints,
\begin{eqnarray}
   ^{(1)}\t{H}{k}{k} = 0 & \Longleftarrow & (\ref{sym1}) \: , \\
   ^{(1)}\t{H}{-}{[ik]} = 0 & \Longleftrightarrow & (\ref{momconstrfirstorder}) \: . \label{symH}
\end{eqnarray}

At first order the relations between the parts of the Weyl tensor and shear tensor, respectively Ricci tensor, become somewhat simpler. With the first--order shear tensor,
\begin{equation}
    \mbox{}^{(1)} \t{\sigma}{i}{j} = \tdot{P}{i}{j} - \frac{1}{3} \dot{P} \t{\delta}{i}{j}\;,
\end{equation}
we find from (\ref{Esigma}),
\begin{equation}
   \mbox{}^{(1)\!} \t{E}{i}{j} = - \mbox{}^{(1)} \tdot{\sigma}{i}{j} - 2H \mbox{}^{(1)}\t{\sigma}{i}{j} \: , \label{Esigmafirstorder}
\end{equation}
respectively $\mbox{}^{(1)\!} \t{E}{-}{ij} = - \mbox{}^{(1)} \tdot{\sigma}{-}{ij}$ with lowered index. With the first--order term for the magnetic part of the Weyl tensor above, its relation to the time--derivative of the 3--Ricci tensor 
simplifies compared with (\ref{randh}),
\begin{equation}
   2 \t{\delta}{-}{m(i} \t{\epsilon}{mkl}{} \, \mbox{}^{(1)\!} \t{H}{-}{j)k \parallel l} = - \frac{1}{a} \mbox{}^{(1)\!} \tdot{R}{-}{ij} \: .
\end{equation}

%------------------------------------------------------------------------%
\subsubsection{General solution for the first--order trace part}

In this section we shall derive a general solution for the trace evolution equation (\ref{rayfirstorder}). The homogeneous Friedmann equation (\ref{friedmann}) determines the scale factor $a(t)$. With this, we separate the time and spatial derivatives and make the ansatz
\begin{equation}
   \t{P}{a}{i}(X,t) = \mbox{}^{0} \t{Q}{a}{i}(X) + q_{1}(t) \mbox{}^{1} \t{Q}{a}{i}(X) + q_{2}(t) \mbox{}^{2} \t{Q}{a}{i}(X) \: ,
\end{equation}
where the time functions $q_{1/2}(t)$ are the two solutions of the linear differential equation
\begin{equation}
   \ddot{q} + 2 \frac{\dot{a}}{a} \dot{q} + \Big( 3 \frac{\ddot{a}}{a} - \Lambda \Big) (q + q(t_0)) = 0 \: . \label{diffeqn}
\end{equation}
Note that the first part in the ansatz reflects the nonvanishing of the initial perturbation field since here we have to take into account initial conditions for the perturbation field as well as its first and second time derivatives. Solving (\ref{tracefirstorder2}) instead of (\ref{rayfirstorder}) we have three solutions in a natural way since there the differential equation is inhomogeneous. (The result is the same as we show in the example in the appendix where we explicitly solve the inhomogeneous equation.)
Note also that if we insert the above ansatz for $\t{P}{a}{i}(X,t)$ into the first-order Raychaudhuri equation (\ref{rayfirstorder}), we get the constraint
\begin{equation*}
   \mbox{}^{0} \t{Q}{a}{i}(X) = q_{1}(t_0) \mbox{}^{1} \t{Q}{a}{i}(X) + q_{2}(t_0) \mbox{}^{2} \t{Q}{a}{i}(X) \; .
\end{equation*}

With the ansatz and its time derivatives we find
\begin{eqnarray}
   \mbox{}^{1} \t{Q}{a}{i} &\!=\!& + \frac{\dot{q}_{2}(t_0) \tddot{P}{a}{i}(t_0) - \ddot{q}_{2}(t_0) \tdot{P}{a}{i}(t_0)}{\ddot{q}_{1}(t_0)\dot{q}_{2}(t_0) - \dot{q}_{1}(t_0)\ddot{q}_{2}(t_0)} \: , \\
   \mbox{}^{2} \t{Q}{a}{i} &\!=\!& - \frac{\dot{q}_{1}(t_0) \tddot{P}{a}{i}(t_0) - \ddot{q}_{1}(t_0) \tdot{P}{a}{i}(t_0)}{\ddot{q}_{1}(t_0)\dot{q}_{2}(t_0) - \dot{q}_{1}(t_0)\ddot{q}_{2}(t_0)} \: , \label{firstorderQ2}
\end{eqnarray}
and
\begin{equation}
   \mbox{}^{0} \t{Q}{a}{i} = \t{P}{a}{i}(t_0) - q_1(t_0) \mbox{}^{1} \t{Q}{a}{i} - q_2(t_0) \mbox{}^{2} \t{Q}{a}{i} \: .
\end{equation}
Hence, the first-order ``peculiar''--coframe takes the following form:
\begin{equation}
   \mbox{}^{(1)}\t{\tilde{\eta}}{a}{i} = \t{\mathring{\eta}}{a}{i} + \big( q_1(t) - q_1(t_0) \big) \mbox{}^{1} \t{Q}{a}{i} + \big( q_2(t) - q_2(t_0) \big) \mbox{}^{2} \t{Q}{a}{i}, \label{firstorderetatilde}
\end{equation}
where $\t{\mathring{\eta}}{a}{i} \equiv \t{\delta}{a}{i} + \t{P}{a}{i}(t_0)$ is the coframe at the initial time.
Finally, we define the peculiar--quantities $\t{u}{a}{i}$ and $\t{w}{a}{i}$ by
\begin{eqnarray}
   \adot{a}{i} &\!=\!& H \a{a}{i} + \t{u}{a}{i} \; , \quad \t{u}{a}{i} \equiv a \tdot{P}{a}{i} \: , \nonumber \\
   \addot{a}{i} &\!=\!&  \frac{\ddot{a}}{a} \a{a}{i} + \t{w}{a}{i} \; , \quad \t{w}{a}{i} \equiv  a \tddot{P}{a}{i}+ 2 \dot{a} \tdot{P}{a}{i} \: . \nonumber
\end{eqnarray}
They are related to the spatial functions $\t{Q}{a}{i}$ by
\begin{eqnarray}
   \t{w}{a}{i} - \Big( 2H + \frac{\ddot{q}_2}{\dot{q}_2} \Big) \t{u}{a}{i} &\!=\!& a \, \frac{\ddot{q}_{1}\dot{q}_{2} - \dot{q}_{1}\ddot{q}_{2}}{\dot{q}_2} \, \mbox{}^{1} \t{Q}{a}{i} \: , \\
      \t{w}{a}{i} - \Big( 2H + \frac{\ddot{q}_1}{\dot{q}_1} \Big) \t{u}{a}{i} &\!=\!& - a \, \frac{\ddot{q}_{1}\dot{q}_{2} - \dot{q}_{1}\ddot{q}_{2}}{\dot{q}_1} \, \mbox{}^{2} \t{Q}{a}{i} \: , \label{firstorderwuQ2}
\end{eqnarray}
so we are able to express the first-order coframe in terms of these quantities.

%------------------------------------------------------------------------%
%------------------------------------------------------------------------%
\section{Relativistic Zel'dovich Approximation (RZA)}
\label{sect:rza}

First, a comment concerning the use of the wording 
``Relativistic Zel'dovich Approximation'' in previous papers is in order,
avoiding from the beginning of this section confusions that may arise 
during the presentation. Previous work, e.g. \cite{kasai}, suggested to use the relativistic deformation field (see (\ref{RZAcoframe}) below) in analogy to 
the Lagrangian deformation gradient of spatial derivatives of the Newtonian trajectory field (\ref{NZA}). While the density field is then calculated through its exact integral
(\ref{generalrho}), evaluated for this deformation field in accord with the Newtonian expression (\ref{NZArho}), the spatial metric, the spatial Ricci 
curvature and other variables are still taken to be those of the strictly linearized case.
In order to explain why our point of view will differ, we recall Zel'dovich's extrapolation idea at the basis of his approximation \cite{zeldovich:fragmentation,zeldovich:fragmentation2,shandarin&zeldovich,zeldovich:rev} in relation to the
exact foundations of this approximation in the framework of a Lagrangian perturbation theory \cite{buchert89,buchert92}. 
Zel'dovich indeed used the exact integral for the density field in the Lagrangian picture of fluid motion, well--studied in the context of an inertial continuum, and adjusted
the coefficient functions in that integral, so that its linearization on a homogeneous--isotropic background cosmology would reproduce the 
result of the Eulerian linear perturbation theory. He considered only the growing mode solution that, asymptotically, supports the 
parallelity condition (\ref{parallelity}). While in the beginning he combined the exact solution for an inertial continuum with the linearized solution of
gravitational instability, it was later confirmed by Doroshkevich et al. \cite{doroshkevichetal} that this ansatz for the trajectory field also
self--consistently solves the divergence equation for the peculiar--field strength. 
Thus, the motivation was born by appealing to the exact solution of the inertial continuum \cite{zeldovichmyshkis}, and Doroshkevich et al. ({\it loc.cit.}) added the decisive consistency test in the framework of self--gravitating motion. However, as we shall explicitly explain below, we can strictly define the extrapolation idea in the framework of the full set of Lagrangian equations for self--gravitating motion, as was done in \cite{buchert89}, and we employ this definition also in the relativistic context. One consequence of our definition is that we shall functionally evaluate all field variables without linearizing the functional expressions.

%------------------------------------------------------------------------%
\subsection{Definition: The ``Relativistic Zel'dovich Approximation'' (RZA)}

Within a fully Lagrangian perturbation approach the 
trajectory field (\ref{NZA}) solves the full Lagrange--Newton--System (\ref{LNS1},\ref{LNS2}) to first order, while it is important to emphasize that this latter system exclusively contains the deformation gradient $f^a_{\;|i}$ as the only dynamical field variable. Hence, this fact suggests that it is possible to consider the first--order solution of
the LNS as an input into other fields that, like the density integral, are just definitions and can be functionally evaluated for any trajectory field.
Given this remark we aimed in the present work to also write down Einstein's equations for only one field variable, the nonintegrable deformation
coefficients $\eta^a_{\;\,i}$, and consider the linearized solution as an input into functionals of these deformation coefficients.

We think that this point of view best reflects Zel'dovich's extrapolation idea. If we would linearize all functionals, e.g. the metric as a quadratic form
of the deformation field, we would just repeat the well--known result of the linearized Einstein equations. If we would keep one nonlinear expression like the density integral, the extrapolation idea is not consistently applied. Note that by taking the exact density integral we make sure that mass is conserved for any given perturbative solution; but also: by taking the quadratic form of the metric we make sure that we measure distances correctly for any given perturbative solution (being important for setting up the light cone structure for a given order of approximation); similar remarks apply for other fields.
We therefore propose the following definition.

\smallskip
\noindent
\underbar{\sl Definition: ``Relativistic Zel'dovich Approximation''}

\smallskip\noindent
{\sl We consider the $9$ functions in the coframe coefficients as the only variables in the full set of ADM equations for the matter model 
``irrotational dust'' within a flow--orthogonal foliation of spacetime. 
We then consider the general linearized solution for these coefficients. 
The approximation ``RZA'' consists in exactly evaluating any other field as a functional of the linearized solution, without performing further approximations
or truncations.}

\smallskip

The following is a restriction that complies with the use of the Newtonian form of Zel'dovich's approximation.
We restrict the general first--order solution (\ref{firstorderetatilde}) to its trace part and subject it to the condition
\begin{equation}
   \mbox{}^{2} \t{Q}{a}{i}(X) = 0 \: , \label{growingmodeonly}
\end{equation}
i.e.~we only consider the growing mode solution. Hence, with (\ref{firstorderwuQ2}) the Zel'dovich restriction for the peculiar--fields reads
\begin{equation}
   \t{w}{a}{i} = \Big( 2H + \frac{\ddot{q}_1}{\dot{q}_1} \Big) \t{u}{a}{i}  \: , \label{RZArestriction1}
\end{equation}
or, in terms of the initial perturbation field via (\ref{firstorderQ2}),
\begin{equation}
   \tddot{P}{a}{i}(X,t_0) = \frac{\ddot{q}_1(t_0)}{\dot{q}_1(t_0)} \tdot{P}{a}{i}(X,t_0) \: . \label{RZArestriction2}
\end{equation}
With this restriction we have
\begin{equation}
   \mbox{}^{1} \t{Q}{a}{i}(X) = \frac{1}{\dot{q}_1(t_0)} \tdot{P}{a}{i}(X,t_0) \: ,
\end{equation}
and we find for the ``peculiar''--coframe,
\begin{equation}
   \mbox{}^{\RZA}\t{\tilde{\eta}}{a}{i}(X,t) = \t{\delta}{a}{i} + \t{P}{a}{i}(X,t_0) + \xi(t) \tdot{P}{a}{i}(X,t_0) \: , \label{RZAcoframe}
\end{equation}
where
\begin{equation}
   \xi(t) \equiv \frac{q_1(t) - q_1(t_0)}{\dot{q}_1(t_0)} \: .
\end{equation}
Remember that the ``peculiar''--coframes were defined by $\t{\tilde{\eta}}{a}{i}(X,t) = \frac{1}{a(t)} \, \a{a}{i}(X,t)$.

Apart from the term that arises because of the nonvanishing initial perturbation, this solution is familiar from the section on Newtonian dynamics above. The corresponding expression for the deviation field $\t{p}{a}{i}$ takes, because of $\tdot{P}{a}{i}(t_0) = \tdot{p}{a}{i}(t_0) - H(t_0) \t{p}{a}{i}(t_0)$, the form:
\begin{eqnarray}
   \mbox{}^{\RZA}\t{\tilde{\eta}}{a}{i} = \t{\delta}{a}{i} + \big( 1 - H(t_0) \big) \, \t{p}{a}{i}(t_0) + \xi(t) \, \tdot{p}{a}{i}(t_0) \: . \nonumber
\end{eqnarray}

We furthermore suggest, and we imply this in our general definition above, to extend the extrapolating approximation RZA to the {\it general} first--order solution, notably including its trace--free part where, this latter,
allows to define a nonlinear approximation for gravitational radiation.

%------------------------------------------------------------------------%
\subsection{Functional evaluation of field variables}

As a consequence of the above definition we are now going to evaluate a number of relevant fields. In this section we write
\begin{equation}
\t{P}{a}{i} \equiv \t{P}{a}{i}(X,t_0) \quad \textrm{and} \quad \tdot{P}{a}{i} \equiv \tdot{P}{a}{i}(X,t_0) \: ,
\end{equation}
because $a(t)$, $\xi(t)$ are the only time--dependent functions.

As said in the above definition of the RZA, we insert the coframe $\mbox{}^{\RZA}\a{a}{i} = a (\t{\mathring{\eta}}{a}{i} + \xi \tdot{P}{a}{i})$, where
\begin{equation}
   \t{\mathring{\eta}}{a}{i} \equiv \t{\delta}{a}{i} + \t{P}{a}{i}(X,t_0)
\end{equation}
is the coframe at some initial time $t_0$, into the exact definitions and equations, as given in section \ref{sect:theory}. The symmetry condition (\ref{LES_E1}), for example, is extrapolated to
\begin{equation}
   \big( 2H\dot{\xi} + \ddot{\xi} \big) \big( \tdot{P}{-}{[ij]} + \delta_{ab} \tdot{P}{a}{[i} \t{P}{b}{j]} \big) = 0 \: .
\end{equation}

The metric for the coframe $\mbox{}^{\RZA}\a{a}{i} = a \, (\, \t{\mathring{\eta}}{a}{i} + \xi \tdot{P}{a}{i}\, )$ takes the following quadratic form:
\begin{equation}
   \mbox{}^{\RZA} \t{g}{-}{ij} = a^2 \t{\delta}{-}{ab} \big[ \t{\mathring{\eta}}{a}{i} \t{\mathring{\eta}}{b}{j} + 2 \xi \t{\mathring{\eta}}{a}{(i} \tdot{P}{b}{j)} + \xi^2 \tdot{P}{a}{i} \tdot{P}{b}{j} \big] \, . \label{RZAmetric}
\end{equation}
Furthermore, we define:
\begin{eqnarray}
   \mbox{}^{(0)\!\!}\t{J}{i}{j} &\!\equiv\!& \frac{1}{6} \t{\epsilon}{-}{abc} \t{\epsilon}{ikl}{} \t{\mathring{\eta}}{a}{j} \t{\mathring{\eta}}{b}{k} \t{\mathring{\eta}}{c}{l}  \: , \nonumber \\
   \mbox{}^{(1)\!\!}\t{J}{i}{j} &\!\equiv\!& \frac{1}{6} \t{\epsilon}{-}{abc} \t{\epsilon}{ikl}{} \tdot{P}{a}{j} \t{\mathring{\eta}}{b}{k} \t{\mathring{\eta}}{c}{l} + \frac{1}{3} \t{\epsilon}{-}{abc} \t{\epsilon}{ikl}{} \t{\mathring{\eta}}{a}{j} \tdot{P}{b}{k} \t{\mathring{\eta}}{c}{l} \: , \nonumber \\
   \mbox{}^{(2)\!\!}\t{J}{i}{j} &\!\equiv\!& \frac{1}{3} \t{\epsilon}{-}{abc} \t{\epsilon}{ikl}{} \tdot{P}{a}{j} \tdot{P}{b}{k} \t{\mathring{\eta}}{c}{l} + \frac{1}{6} \t{\epsilon}{-}{abc} \t{\epsilon}{ikl}{} \t{\mathring{\eta}}{a}{j} \tdot{P}{b}{k} \tdot{P}{c}{l} \: , \nonumber \\
   \mbox{}^{(3)\!\!}\t{J}{i}{j} &\!\equiv\!& \frac{1}{6} \t{\epsilon}{-}{abc} \t{\epsilon}{ikl}{} \tdot{P}{a}{j} \tdot{P}{b}{k} \tdot{P}{c}{l} \: . \nonumber
\end{eqnarray}

With the help of these definitions the transformation determinant $\mbox{}^{\RZA\!} J$ reads:
\begin{equation}
   \mbox{}^{\RZA\!} J = a^3 \big( \t{J}{-}{0} + \xi \t{J}{-}{1} + \xi^2 \t{J}{-}{2} + \xi^3 \t{J}{-}{3} \big) \: ,
\end{equation}
where $\t{J}{-}{n} \equiv \mbox{}^{(n)\!\!}\t{J}{k}{k}$. In the homogeneous case, i.e.~for vanishing initial perturbations, we find $\mbox{}^{H\!\!}\t{J}{-}{0} = 1$ whereas $\mbox{}^{H\!\!}\t{J}{-}{1}$--$\mbox{}^{H\!\!}\t{J}{-}{3}$ reduce to the three scalar invariants for $\tdot{P}{a}{i}$. Hence, the nonlinearly evolved density can be calculated through its exact integral,
\begin{equation}
\label{RZArho}
   \mbox{}^{\RZA\!}\varrho = \frac{\mathring{\rho} \mathring{J}}{\mbox{}^{\RZA\!}J} = \frac{\varrho_H(t)}{\varrho_H(t_0)} \mathring{\varrho}(\vec{X}) \frac{\mathring{J}}{\mbox{}^{\RZA\!} \tilde{J}} \: ,
\end{equation}
where ${\mbox{}^{\RZA\!} \tilde{J}} \equiv \det (\t{\tilde{\eta}}{a}{i}) = \t{J}{-}{0} + \xi \t{J}{-}{1} + \xi^2 \t{J}{-}{2} + \xi^3 \t{J}{-}{3}$ is evaluated for the ``peculiar''--coframe field (\ref{RZAcoframe}) and $\mathring{J} \equiv J(X,t_0)$ is the transformation determinant at some initial time.

Then the orthonormal vector basis $\e{a}{i}$ in the RZA--picture is given by
\begin{equation}
   \mbox{}^{\RZA} \e{a}{i} = \frac{1}{a} \, \Big[ \frac{1}{2\tilde{J}} \, \t{\epsilon}{-}{abc} \t{\epsilon}{ikl}{} \big( \t{\mathring{\eta}}{b}{k} \t{\mathring{\eta}}{c}{l} + 2 \xi \t{\mathring{\eta}}{b}{k} \tdot{P}{c}{l} + \xi^2 \tdot{P}{b}{k} \tdot{P}{c}{l} \big) \Big] \: , \nonumber
\end{equation}
so the orthonormality relation for $\a{a}{i}$ and $\e{a}{i}$ in the RZA becomes (here and in the following the summation always runs from $n=0\dots3\,$):
\begin{equation}
   \mbox{}^{\RZA} \e{a}{i} \;  \mbox{}^{\RZA}\a{a}{j} = 3 \, \frac{\sum_n \xi^n \;\mbox{}^{(n)\!}\t{J}{i}{j}}{\sum_n \xi^n \t{J}{-}{n}} \nonumber
\end{equation}
with trace $\mbox{}^{\RZA}  \e{a}{k} \; \mbox{}^{\RZA}\a{a}{k} = 3$, as expected.

To be able to write the RZA expansion tensor in a similarly short form we define the second set of auxiliary quantities, i.e.
\begin{eqnarray}
   \mbox{}^{(0)\!\!}\t{I}{i}{j} &\!\equiv\!& \frac{1}{6} \t{\epsilon}{-}{abc} \t{\epsilon}{ikl}{} \t{\mathring{\eta}}{a}{j} \t{\mathring{\eta}}{b}{k} \t{\mathring{\eta}}{c}{l}  \: , \nonumber \\
   \mbox{}^{(1)\!\!}\t{I}{i}{j} &\!\equiv\!& \frac{1}{2} \t{\epsilon}{-}{abc} \t{\epsilon}{ikl}{} \tdot{P}{a}{j} \t{\mathring{\eta}}{b}{k} \t{\mathring{\eta}}{c}{l}  \: , \nonumber \\
   \mbox{}^{(2)\!\!}\t{I}{i}{j} &\!\equiv\!& \frac{1}{2} \t{\epsilon}{-}{abc} \t{\epsilon}{ikl}{} \tdot{P}{a}{j} \tdot{P}{b}{k} \t{\mathring{\eta}}{c}{l}  \: , \nonumber \\
   \mbox{}^{(3)\!\!}\t{I}{i}{j} &\!\equiv\!& \frac{1}{6} \t{\epsilon}{-}{abc} \t{\epsilon}{ikl}{} \tdot{P}{a}{j} \tdot{P}{b}{k} \tdot{P}{c}{l} \: . \nonumber
\end{eqnarray}
Of course $\mbox{}^{(0)\!\!}\t{I}{i}{j} = \mbox{}^{(0)\!\!}\t{J}{i}{j}$ as well as $\mbox{}^{(3)\!\!}\t{I}{i}{j} = \mbox{}^{(3)\!\!}\t{J}{i}{j}$ and their traces are identical to those of the $\t{J}{i}{j}$'s.

Hence, the RZA expansion tensor takes the form:
\begin{equation}
   \mbox{}^{\RZA} \t{\Theta}{i}{j} = 3H \cdot \frac{\sum_n \xi^n \;\mbox{}^{(n)\!}\t{J}{i}{j}}{\sum_n \xi^n \t{J}{-}{n}} + \frac{\dot{\xi}}{\xi} \cdot \frac{\sum_n n \, \xi^n \;\mbox{}^{(n)\!}\t{I}{i}{j}}{\sum_n \xi^n \t{J}{-}{n}} \: .
\end{equation}
We evaluate the mixed components of the tensor, since in this form the corresponding Newtonian expressions are easily recovered. The expansion scalar, for example, is
\begin{equation}
   \mbox{}^{\RZA} \Theta = 3H + \frac{\dot{\xi}}{\xi} \cdot \frac{\xi \t{J}{-}{1} + 2 \, \xi^2 \t{J}{-}{2} + 3 \, \xi^3 \t{J}{-}{3}}{\t{J}{-}{0} + \xi \t{J}{-}{1} + \xi^2 \t{J}{-}{2} + \xi^3 \t{J}{-}{3}} \: .
\end{equation}

Now we give the expressions for the parts of the Weyl tensor in the RZA. The electric part reads:
\begin{eqnarray}
   \mbox{}^{\RZA\!} \t{E}{i}{j} = &\!\!-\!\!& 3\frac{\ddot{a}}{a} \cdot \frac{\sum_n \xi^n \;\mbox{}^{(n)\!}\t{J}{i}{j}}{\sum_n \xi^n \t{J}{-}{n}} - \Big( \frac{\ddot{\xi}}{\xi} + 2 H \frac{\dot{\xi}}{\xi} \Big) \frac{\sum_n n \, \xi^n \;\mbox{}^{(n)\!}\t{I}{i}{j}}{\sum_n \xi^n \t{J}{-}{n}} \nonumber \\
    &\!\!-\!\!& \frac{1}{3} \Big( 4 \pi G \frac{\mathring{\varrho}\mathring{J}}{\mbox{}^{\RZA\!}J} -  \Lambda \Big) \t{\delta}{i}{j} \: .
\end{eqnarray}
The relation to the mixed shear tensor components reads:
\begin{eqnarray}
   \mbox{}^{\RZA\!} \t{E}{i}{j} = &\!-\!& \mbox{}^{\RZA} \tdot{\sigma}{i}{j} - 2 H \;\mbox{}^{\RZA} \t{\sigma}{i}{j} - \frac{2}{3} \frac{\dot{\xi}}{\xi} \cdot \frac{\sum_n n \xi^n \t{J}{-}{n}}{\sum_n \xi^n \t{J}{-}{n}} \cdot \mbox{}^{\RZA} \t{\sigma}{i}{j} \nonumber \\
   &\!-\!& \mbox{}^{\RZA} \t{\sigma}{i}{k} \;\mbox{}^{\RZA} \t{\sigma}{k}{j} + \frac{1}{3} \;\mbox{}^{\RZA} \t{\sigma}{k}{l} \;\mbox{}^{\RZA} \t{\sigma}{l}{k} \t{\delta}{i}{j} \;.
\end{eqnarray}
(This should be compared with the somewhat simpler form (\ref{Esigmafirstorder}) in the first--order scheme, and with the general relation (\ref{Esigma}).)

The magnetic part takes the following form:
\begin{equation}
   \mbox{}^{\RZA\!} \t{H}{i}{j} = - \frac{\dot{\xi}}{a} \t{\epsilon}{ikl}{} \cdot \frac{\tdot{P}{-}{kj \parallel l} + \t{\delta}{-}{ab} \big( \tdot{P}{a}{j} \t{P}{b}{k} + \xi \tdot{P}{a}{j} \tdot{P}{b}{k} \big)_{\parallel l}}{\sum_n \xi^n \t{J}{-}{n}} \, ,
\end{equation}
where $\parallel$ here denotes the covariant derivative with respect to the RZA--metric (\ref{RZAmetric}).

Finally, we express the Riemann and Ricci curvature tensors functionally in terms of the RZA--deformation. To keep the equations short, we define (analogous to the  functionals $\mbox{}^{(0)\!\!}\t{J}{i}{j} - \mbox{}^{(3)\!\!}\t{J}{i}{j}$ above) the quantities
\begin{eqnarray}
   \mbox{}^{(0)\!\!}\t{\tilde{R}}{i}{jkl} &\!\equiv\!& \frac{1}{6} \t{\epsilon}{-}{abc} \t{\epsilon}{imn}{} \t{\mathring{\eta}}{a}{j\parallel k \parallel l} \t{\mathring{\eta}}{b}{m} \t{\mathring{\eta}}{c}{n}  \: , \nonumber \\
   \mbox{}^{(1)\!\!}\t{\tilde{R}}{i}{jkl} &\!\equiv\!& \frac{1}{6} \t{\epsilon}{-}{abc} \t{\epsilon}{imn}{} \tdot{P}{a}{j\parallel k \parallel l} \t{\mathring{\eta}}{b}{m} \t{\mathring{\eta}}{c}{n} \nonumber \\
   &\!+\!& \frac{1}{3} \t{\epsilon}{-}{abc} \t{\epsilon}{imn}{} \t{\mathring{\eta}}{a}{j\parallel k \parallel l} \tdot{P}{b}{m} \t{\mathring{\eta}}{c}{n} \: , \nonumber \\
   \mbox{}^{(2)\!\!}\t{\tilde{R}}{i}{jkl} &\!\equiv\!& \frac{1}{3} \t{\epsilon}{-}{abc} \t{\epsilon}{imn}{} \tdot{P}{a}{j\parallel k \parallel l} \tdot{P}{b}{m} \t{\mathring{\eta}}{c}{n} \nonumber \\
   &\!+\!& \frac{1}{6} \t{\epsilon}{-}{abc} \t{\epsilon}{imn}{} \t{\mathring{\eta}}{a}{j\parallel k \parallel l} \tdot{P}{b}{m} \tdot{P}{c}{n} \: , \nonumber \\
   \mbox{}^{(3)\!\!}\t{\tilde{R}}{i}{jkl} &\!\equiv\!& \frac{1}{6} \t{\epsilon}{-}{abc} \t{\epsilon}{imn}{} \tdot{P}{a}{j\parallel k \parallel l} \tdot{P}{b}{m} \tdot{P}{c}{n} \: . \nonumber
\end{eqnarray}
The curvature tensor then is
\begin{equation}
   \mbox{}^{\RZA\!} \t{R}{i}{jkl} = 3 \cdot \frac{\sum_n \xi^n \big( \mbox{}^{(n)\!} \t{\tilde{R}}{i}{jkl} - \mbox{}^{(n)\!} \t{\tilde{R}}{i}{jlk} \big)}{\sum_n \xi^n \t{J}{-}{n}} \;,\label{RZARiemann}
\end{equation}
and the Ricci tensor takes the form
\begin{equation}
   \mbox{}^{\RZA\!} \t{R}{-}{ij} = 3 \cdot \frac{\sum_n \xi^n \big( \mbox{}^{(n)\!} \t{\tilde{R}}{k}{ikj} - \mbox{}^{(n)\!} \t{\tilde{R}}{k}{ijk} \big)}{\sum_n \xi^n \t{J}{-}{n}} \;.\label{RZARicci}
\end{equation}
Of course one can express the curvature quantities solely in terms of the RZA--deformation, since we used the covariant derivative with respect to the RZA--metric (\ref{RZAmetric}), but the so--found equations are long and hard to read.

%------------------------------------------------------------------------%

\section{Discussion and Conclusions}
\label{sect:conclusions}

In this section we put some aspects into perspective that were discussed in the context of the Newtonian ``Zel'dovich approximation'', and we summarize the main points of this paper. 

%------------------------------------------------------------------------%
\subsection{Possible tests of the extrapolation} 

Applications of the presented approximate theory leave in suspense the unknown quality of the extrapolation done. Of course, we may blindly accept the resulting approximate solutions driven by the belief that the corresponding Newtonian model is extremely successful in comparison with N--body simulations of the full problem. However, it is in order to point out that a number of self--consistency tests are possible and should be done. While this scheme can predict and describe effects beyond the known highly symmetric solutions of general relativity, it is necessary to conduct additional tests. Such tests are best performed in the context of the envisaged applications, and we shall come back to them in forthcoming papers. Recall that Doroshkevich et al. \cite{doroshkevichetal} provided such a test for the Newtonian form of Zel'dovich's approximation by considering quantitatively the difference between the density calculated from the exact integral of the continuity equation and the density calculated from the field equation, i.e. from the divergence of the peculiar--gravitational field strength. While at first order both expressions agree by construction, the error was reported to be of second and higher order \cite{doroshkevichetal}. While qualitatively this is obvious, the error was calculated and used to estimate the quantitative validity of the approximation in certain regimes.
In this spirit we can also compare resulting nonlinear expressions and conduct consistency tests. As an example we note that e.g. the scalar curvature can be determined from the Hamilton constraint (\ref{hamconstraint}) through the kinematical invariants and the density in the RZA, and alternatively from the RZA metric by explicitly calculating the trace of its Ricci tensor (\ref{R}). The resulting expressions agree to first order and the error is of higher order and may be quantitatively controlled in the context of a given application. We shall come back to these different curvature expressions in forthcoming papers.

%------------------------------------------------------------------------%
\subsection{Singularities} 

Continua made of {\it dust} are bound to develop singularities in the course of evolution,
resulting in {\it caustics}, i.e. loci of formally infinite density. This is a consequence of neglecting physical effects like velocity dispersion
or vorticity that could regularize singularities (for the Newtonian theory see \cite{buchertdominguez:adhesive}).
In general relativity this situation corresponds to the intersection of world--lines and, thus, to the failure of defining a congruence of world--lines, together 
with the possibility of simultaneously developing singularities in geometrical fields.

The appropriate mathematical framework in which caustics can be described and 
classified is {\it catastrophe theory} \cite{thom:book}, further developed in the framework of the
{\it Lagrange--singularity theory} especially by Vladimir Arnol'd and collaborators. 
Caustics are defined as images of a critical set on a
{\it Lagrangian submanifold} under a projection map.
The stable singularities of such {\it Lagrangian mappings} can be classified
into a finite set of topological structures ({\it germs}), their number depending on
the number of dimensions of the Lagrangian submanifold. 
A classification into a {\it finite} number of elements
is only possible for manifolds with dimension $\le 5$.
Alternatively, the {\it Lagrange--singularities} of a family of
world--lines of fluid elements can be described in terms of {\it Legendre--singularities}
of the wavefronts (if they exist) as the dual description of the continuum's evolution
({\it cf.} the small selection of papers by 
Arnol'd  \cite{arnoldrussian}, \cite{arnold1}, \cite{arnold2}, 
the books \cite{arnold:book} and the collection of papers \cite{arnold:editor}).

The singularities developing in an irrotational Newtonian continuum moving
under inertia have been analyzed in full detail with nice hand drawings by 
Vladimir Arnol'd \cite{arnoldrussian},
and a formal relation of this mapping to the (Newtonian) ``Zel'dovich approximation'' 
has been established and analyzed in detail for caustics in two spatial dimensions \cite{arnoldshandarinzeldovich}. 
In the {\it pancake picture} developed at the time
one considers images of singularities as the {\it local} building blocks
of the large--scale structure in the Universe. The geometry of these structures may
differ for different realizations of the model, but the local
morphology of structures (so--called {\it unfoldings} 
around singularities and their evolution (so--called {\it metamorphoses}) 
is completely made up from $12$ (topologically classified) elements in a four--dimensional continuum \cite{arnoldrussian}.

Specifically, in Newtonian theory, we may define a flow field ${\bf x} = {\bf f}({\bf X},t)$ by a time--dependent diffeomorphism that sends 
initial (Lagrangian) positions $\bf X$ of fluid elements to their Eulerian positions $\bf x$ (embedded into Euclidian space) at time $t$.
In order to apply the classification scheme of the Lagrange singularity theory to the motion of a Newtonian continuum, the
key--property that has to be required for $\bf f$ is that it 
can be written as a family of {\it gradient mappings},
which form an important class of {\it Lagrangian mappings}.
This is for example easily possible, if the flow is {\it irrotational with respect to the Lagrangian frame}, i.e., there exists a potential
$\Psi$ for which ${\bf f} =:\boldsymbol{\nabla}_{\bf X} \Psi$, where $\boldsymbol{\nabla}_{\bf X}$ denotes derivative with respect to Lagrangian coordinates. 
We may then define the one--parameter family (parametrized by the time $t$) of Lagrangian mappings ${\pi}_t$:
\begin{equation}
{\pi}_t: {\mathbb R}^3 \rightarrow {\mathbb R}^3\;\;\;;\;\;\;
{\bf X} \mapsto {\bf x} = \boldsymbol{\nabla}_{\bf X} \Psi ({\bf X};t) \;\;,
\end{equation}
where in this case the set $\lbrace \bf X, \bf x=\bf f(\bf
X,t)\rbrace$ forms a {\it Lagrangian submanifold} of ${\mathbb R}^6 = \lbrace
\bf X, \bf x \rbrace$.
Then, for each fixed $t$, the (non--degenerate closed) two--form $\sum_i
{\bf d}x^i \wedge {\bf d}X^i$ vanishes on the Lagrangian submanifold
($X^i$ are local coordinates on this manifold). Note that the requirement of 
irrotationality of the flow field with respect to Lagrangian coordinates is in general
much more restrictive than the requirement of irrotationality with respect to Eulerian coordinates.
For rotational flows, the Lagrange--singularity theory has been extended by 
Bruce and collaborators \cite{bruce}, specifically investigated for
solenoidal velocity fields of an inertial continuum. They found that the most ubiquitous
{\it pancake}--singularities ($A_3$ in Arnol'd's classification)
remain stable. However, the singularities associated with {\it umbilics}
in the potential case have to be removed from the list of {\it generic} (structurally stable) patterns in the vortical case.

The possibility of writing the flow field $\bf f$ in the form of a 
{\it gradient mapping} can be 
demonstrated for some important subclasses of irrotational Lagrangian perturbation 
solutions: the flow fields develop {\it Lagrange--singularites} in the case of first--order solutions \cite{buchert92},
and for a large class of second--order solutions \cite{buchertehlers93}. 
The third-- and fourth--order contributions destroy this possibility even for restricted classes of initial conditions \cite{buchert94,rampfbuchert}.

In the case of self--gravitating continua, a proof of the Lagrangian property meets the problem that the
velocity field {\it and} the acceleration field may become multi--valued
simultaneously (\cite{arnold3} \S 6, footnote 1). Nonperturbatively the situation is worse, since 
the gravitational field strength does not remain finite at caustics, as would be suggested by
Lagrangian perturbation solutions to any order \cite{buchert:integral1,buchert:integral2}.
Indeed, a consequence of the action of self--gravity is the development of a hierarchy of nested caustics, which originate simultaneously,
with an ever increasing number of streams  (see \cite{buchertehlers93} for an example in second--order perturbation solutions).

In general relativity the problems described above are more involved.
One open problem is the dynamical justification of the continuation of
solutions across caustics, where the multi--stream flow is required to satisfy the
field equations. A mathematically well--defined concept of bifurcating {\it dust}
appears to be a difficult problem; it is evident that also the shapes of caustic
surfaces may not necessarily permit a morphological classification
in the framework of the Lagrange--singularity theory. Clarke and O'Donnel \cite{clarkeodonnell}
succeeded in showing the self--consistency of an extension of spacetime through a dust caustic.
Other authors address the problem  in the spherically symmetric case \cite{frauendienerklein}.
In cosmology the singularity problem is mostly addressed for the asymptotic past of solutions;  
for a study of the asymptotic structure of cosmological singularities see, e.g., \cite{Lim1,Lim2}.
Of course there is a substantial literature related to singularities in general relativity (see, e.g., 
\cite{kriele,nolan,clarkeschmidt}, and references therein, as well as the book by Clarke \cite{clarke:book}.

The model proposed in this paper contributes to this discussion. We can explicitly study those fields that would feature singularities
in the relativistic Zel'dovich approximation. Since caustics correspond, in the Newtonian model, to degeneracies of the Jacobian, $J = \det (f^i_{\;|k}) = 0$, we have to look
at degeneracies of the corresponding relativistic field (denoted by the same letter in the present paper), $J = \det (\eta^a_{\;k}) = 0$, in the local exact basis ${\bf d}X^k$. 
From this it is already evident
that the nonintegrability of Cartan's deformation field  -- in general -- destroys the possibility of defining a family of gradient mappings, and a morphological classification
in the classical framework is not straightforward. With the reasonable assumption that local deformations remain finite we conclude that the metric, as a quadratic form of the deformations, remains finite at caustics. However, as the explicit formulae in Appendix C show, almost all relevant fields will degenerate simultaneously, since they are weighted by  $\mbox{}^{\RZA\!} J$ (while the remaining terms remain finite for finite deformations). Note here that the verticality (with respect to Eulerian coordinates) of the Newtonian velocity gradient, which is indicative for the occurrence of a singularity, corresponds to the relativistic field $\Theta^a_{\;b}$ that becomes the mixed--index object $\Theta^i_{\;j} = e_a^{\;i}\eta^b_{\;j} \Theta^a_{\;b}$ in the exact basis, {\it cf.} Eq.~(\ref{mixedTheta}). The explicit expressions in the RZA model imply that spatial but also spacetime curvature terms become singular at caustics.
Due to the form of the general expressions, however, we expect these degeneracies to appear also in general situations.

%------------------------------------------------------------------------%
\subsection{Concluding Remarks}

Following the systematic derivation of Zel'dovich's approximation in the Lagrangian framework of the Newtonian equations we have formulated the 
Einstein equations for the matter model `irrotational dust' in terms of a single dynamical variable. The nine coframe coefficients of Cartan's deformation
one--forms replace the integrable Newtonian deformation gradient deriving from three vector components. We discussed the resulting system using
different representations, and we derived the general first--order solution for the coframe functions (the deformation field). 
We then gave a definition of a nonperturbative approximation scheme that proposes to functionally evaluate dynamical field variables in terms of the perturbed deformation field. 
The success of the corresponding Newtonian approximation gives substantial motivation for this generalization to relativistic cosmology. 

There are a number of aspects that we consider useful. First, using the proposed equations we can easily translate ``Newtonian knowledge'' to the relativistic stage. This is especially due to the formally close correspondence of the `electric part of the LES' to the LNS of Newtonian theory.
Second, we can employ explicit forms of nonperturbative expressions for field variables that just depend on initial data and known
time--dependent functions. We so are able to attack highly nonlinear problems in relativistic cosmology. For example, the approximate quadratic form
of the metric can be used to realistically evaluate distance expressions in inhomogeneous structure distributions; the explicit structure of the light cone allows the study of the influence of generic inhomogeneities. The same is true for the Ricci curvature and the Weyl curvature with its parts.     
As a consequence this approximation allows to investigate many problems beyond the Newtonian approximation such as light propagation and gravitational radiation. 

In forthcoming work we extend this model by employing exact integral properties of Einstein's equations \cite{buchert:onaverage1}. The combination of a generic model for inhomogeneous deformations with exact integral properties has led to a powerful model in the Newtonian approximation \cite{buchert:bks,kerscher:abundance}, and so we shall investigate the corresponding relativistic problem in order to quantify the influence of inhomogeneities on average properties of the Universe (the `backreaction problem' \cite{darkenergy:review,buchert:focus,ellisFOCUS}). Furthermore, we shall give the general perturbation and solution schemes for the Lagrange--Einstein--System including examples of higher--order Lagrangian perturbation solutions. Nonperturbative investigations of light propagation in inhomogeneous models as well as nonlinear aspects of gravitational radiation are also envisaged
in future applications. 

A further, more challenging but possible, application of the presented formalism would employ a self--consistently evolving background rather than a fixed FLRW background as in our examples. A background including backreaction effects could be determined by the exact average properties of an inhomogeneous universe model \cite{generalbackground}. Including pressure by translating Newtonian results is possible \cite{adlerbuchert,buchertdominguez:adhesive}. Furthermore, a nonvanishing shift vector field together with a 
non--constant lapse function could be included and herewith the Lagrangian 
condition extended, all providing more general frameworks in the spirit of this work.

%------------------------------------------------------------------------%
%------------------------------------------------------------------------%
\section*{Acknowledgements}

{\small This work has a long history. It started with the analogy of the Lagrange--Newton--System to what is called `electric part of the Lagrange--Einstein--System' in this paper, developed in 1996 during a visit of TB to the Albert--Einstein--Institut in Potsdam, whose hospitality and support is acknowledged. Thereafter, the formalism and its application to the relativistic Zel'dovich approximation were the subject of two diploma thesis works \cite{lante,ostermann}, supported by ``Sonderforschungsbereich 375 f\"ur Astro--Teilchenphysik der Deutschen Forschungsgemeinschaft". During this early stage of the work we have profited from interesting discussions with J\"urgen Ehlers, Toshi Futamase, Masumi Kasai, Martin Kerscher, Stephan Lante, Sabino Matarrese, Jens Schmalzing, Christian Sicka, and Herbert Wagner. We are especially grateful to Herbert Wagner for his continuous 
encouragement. Valuable comments on the final manuscript have been given by Alexandre Alles and Alexander Wiegand. We thank an anonymous referee for pointing to the singularity issue.
This work was supported by ``F\'ed\'eration de Physique Andr\'e--Marie Amp\`ere, Lyon'', and was conducted within the ``Lyon Institute of Origins'' under grant 
ANR-10-LABX-66.}

%------------------------------------------------------------------------%
%------------------------------------------------------------------------%
\renewcommand{\theequation}{A.\arabic{equation}}
\setcounter{equation}{0} 
\section*{APPENDIX A: Formulation using Differential Forms}

\subsection*{Newtonian equations}

An alternative, somewhat simpler form of the Lagrange--Newton--System of equations can be reproduced by introducing differential forms. Using the 
spatial exterior derivative operator $\textbf{d}$, acting on functions and forms, and the spatial exterior product, the LNS takes the form \cite{ehlersbuchert}:
\begin{eqnarray}
   \delta_{ij} \textbf{d} \ddot{f}^i \wedge \textbf{d} f^j &\!=\!& {\bf 0} \, ; \\
   \frac{1}{2} \epsilon_{ijk} \textbf{d} \ddot{f}^i \wedge \textbf{d} f^j \wedge \textbf{d} f^k &\!=\!& \left( \Lambda- 4 \pi G \varrho \right) \textbf{d}^3 f \, ,
\end{eqnarray}
where the density is given by the integral $\varrho = \mathring{\varrho}{J}^{-1}$, and $\textbf{d}^3 f = J \textbf{d}^3 X$. Defining the three differential one--forms of the tidal tensor $\boldsymbol{\cal E}^i = {\cal E}^i_{\;|j} \textbf{d}X^j$, the LNS assumes instead the compact form:
\begin{eqnarray}
   \delta_{ij} \boldsymbol{\cal E}^i \wedge \textbf{d} f^j &\!=\!& {\bf 0} \, ;\\
   \epsilon_{ijk} \boldsymbol{\cal E}^i \wedge \textbf{d} f^j \wedge \textbf{d} f^k &\!=\!& {\bf 0} \, ,
\end{eqnarray}
where
\begin{equation}
   \boldsymbol{\cal E}^i = \textbf{d} \ddot{f}^i -\frac{1}{3} \left( \Lambda - 4 \pi G \varrho \right) \textbf{d} f^i \, .
\end{equation}

\subsection*{Einstein equations}

We introduced Cartan's coframes $\ba{a} = \a{a}{i} \, \textbf{d}X^i$, one--forms that define a noncoordinate basis of three--dimensional space. We may call them \emph{spatial deformation one--forms}. The metric form is the canonical quadratic form
\begin{equation}
   \bg = \t{\delta}{-}{ab} \ba{a} \otimes \ba{b} \: .
\end{equation}
Noncoordinate indices are raised and lowered by $\t{\delta}{-}{ab}$.

The expansion one--form is then defined by the parallel transport equation, i.e. 
\begin{equation}
   \boldsymbol{\Theta}^a = \t{\Theta}{a}{b} \ba{b} = \dot{\boldsymbol{\eta}}^a = \adot{a}{i} \, \textbf{d}X^i \, .
\end{equation}
Spatial derivatives that take into account the varying geometry are evaluated, for a symmetric connection, by  
Cartan's connection one--forms,
\begin{equation}
   \t{\boldsymbol{\omega}}{a}{b} = \t{\gamma}{a}{cb} \ba{c}
\end{equation}
together with the curvature two--form
\begin{equation}
   \t{\boldsymbol{\Omega}}{a}{b} = \frac{1}{2}\t{\CR}{a}{bcd} \ba{c} \wedge \ba{d} \, .
\end{equation}
These objects are defined by the (spatial) Cartan structure equations:
\begin{eqnarray}
   \boldsymbol{\omega}_{ab} + \boldsymbol{\omega}_{ba} &\!=\!& {\bf 0} \;;\\
   \textbf{d} \ba{a} + \boldsymbol{\omega}^a_{\;\;b} \wedge \ba{b} &\!=\!& \textbf{T}^a \;;\\
   \textbf{d} \boldsymbol{\omega}^a_{\;\;b} + \boldsymbol{\omega}^a_{\;\;c} \wedge \boldsymbol{\omega}^c_{\;\;b} &\!=\!& \boldsymbol{\Omega}^a_{\;\;b} \; ,
\end{eqnarray}
together with the integrability conditions $\textbf{d}\textbf{d}\ba{a} = \textbf{0}$ and $\textbf{d}\textbf{d}  \t{\boldsymbol{\omega}}{a}{b} = \textbf{0}$,
where $\textbf{T}^a = \frac{1}{2}\t{T}{a}{bc} \ba{b} \wedge \ba{c}$ is the torsion two--form. We set the torsion to zero in this work. Hence, both $\t{\boldsymbol{\omega}}{a}{b}$ and $\t{\boldsymbol{\Omega}}{a}{b}$ can, in principle, be expressed solely in terms of the deformation one--forms. 

We may define the total exterior derivative of a 
tensor--valued differential form $\boldsymbol{\Phi}$ by
\begin{equation}
   \left( \textbf{D} \Phi \right)^a_{\;\;b} = \textbf{d} \Phi^a_{\;\;b} + \boldsymbol{\omega}^a_{\;\;c} \wedge \Phi^c_{\;\;b} - \boldsymbol{\omega}^c_{\;\;b} \wedge \Phi^a_{\;\;c}\;,
\end{equation}
which is the natural generalization of the covariant derivative when working in a noncoordinate basis. 
With this definition we have:
\begin{equation}
   \textbf{D} \ba{a} = \textbf{T}^a \quad,\quad  \textbf{D} \textbf{T}^a = \boldsymbol{\Omega}^a_{\;\;b} \wedge \ba{b}\quad \textrm{and} \quad \textbf{D} \boldsymbol{\Omega}^a_{\;\;b} = {\bf 0} \; ,
\end{equation}
where the two last equations represent the two Bianchi identities.

The invariant volume element in the noncoordinate basis is
\begin{equation}
   \ba{1} \wedge \ba{2} \wedge \ba{3} = J \textbf{d}X^1 \wedge \textbf{d}X^2 \wedge \textbf{d}X^3 = J \textbf{d}^3 X \, .
\end{equation}
The Lagrange--Einstein--System takes the form
\begin{eqnarray}
   \delta_{ab} \ddot{\boldsymbol{\eta}}^a \wedge \ba{b} &\!=\!& {\bf 0} \, , \label{LESoneform} \\
  \frac{1}{2} \epsilon_{abc} \ddot{\boldsymbol{\eta}}^a \wedge \ba{b} \wedge \ba{c} &\!=\!& \left( \Lambda J - 4 \pi G \mathring{\varrho}  \right) \textbf{d}^3 X \, , \label{LEStwoform} \\
   \t{\epsilon}{-}{abc} \textbf{D} \dot{\boldsymbol{\eta}}^a \wedge \ba{b} &\!=\!& {\bf 0} \;,\label{LESthreeform}
\end{eqnarray}
and
\begin{eqnarray}
   \frac{1}{2} \epsilon_{bcd} \ddot{\boldsymbol{\eta}}^a \wedge \ba{c} \wedge \ba{d} &\!+\!& \epsilon_{bcd} \dot{\boldsymbol{\eta}}^a \wedge \dot{\boldsymbol{\eta}}^c \wedge \ba{d} \\
   &\!=\!& \left[ \left( 4 \pi G \mathring{\varrho} + \Lambda J \right) \delta^{\;a}_b - J \CR^{\;\,a}_{b} \right] \textbf{d}^3 X \; , \nonumber
\end{eqnarray}
where the 3--Ricci tensor can be expressed via the curvature two--form by
\begin{equation}
   \CR^a_{\;\; d} \ba{d} \wedge \ba{b} \wedge \ba{c} = \delta^{db} \boldsymbol{\Omega}^a_{\;\; d} \wedge \ba{c} - \delta^{dc} \boldsymbol{\Omega}^a_{\;\; d} \wedge \ba{b} \; .
\end{equation}

Let $\Phi = \Phi_{i_1 ... i_r} \textbf{d} x^{i_1} \wedge ... \wedge \textbf{d} x^{i_r}$ be an $r$--form in a three--dimensional manifold and $g = \det (g_{ij})$ the determinant of the metric. Then the \emph{duality operator}, or Hodge star operator, is defined by
\begin{equation}
   \ast \Phi = \frac{\sqrt{\mid g \mid}}{(3-r)!} \Phi_{i_1 ... i_r} \epsilon^{i_1 ... i_r}_{\qquad \;\; i_{r+1} ... i_3} \textbf{d} x^{i_{r+1}} \wedge ... \wedge \textbf{d} x^{i_3} \; . \label{hodge}
\end{equation} 
In particular, $\ast 1$ is the invariant volume element because of
\begin{eqnarray}
   \ast 1 & = & \frac{\sqrt{\mid g \mid}}{3!} \epsilon_{i_1 i_2 i_3} \textbf{d} x^{i_1} \wedge \textbf{d} x^{i_2} \wedge \textbf{d} x^{i_3} \nonumber \\
   & = & \sqrt{\mid g \mid} \textbf{d} x^1 \wedge \textbf{d} x^2 \wedge \textbf{d} x^3 \; .
\end{eqnarray}

With the help of (\ref{hodge}) we can also introduce one--forms of the electric (tidal) and magnetic parts of the Weyl two--form, which are
\begin{eqnarray}
   \textbf{E}^a &\!=\!& - \ddot{\boldsymbol{\eta}}^a + \frac{1}{3} \left( \Lambda - 4 \pi G \varrho \right) \ba{a} \, , \label{Eform} \\
   \textbf{H}^a &\!=\!& \ast \left( \t{\delta}{-}{bc} \textbf{D} \dot{\boldsymbol{\eta}}^c \wedge \ba{a} \right) \ba{b} \, . \label{Hform}
\end{eqnarray}
The differential form counterparts to the equations (\ref{Esym}) -- (\ref{Htrace}) then read:
\begin{eqnarray}
   \delta_{ab} \textbf{E}^a \wedge \ba{b} &\!=\!& {\bf 0} \, ;\label{E1form}\\
   \epsilon_{abc} \textbf{E}^a \wedge \ba{b} \wedge \ba{c} &\!=\!& {\bf 0} \, ;\label{E2form} \\
   \delta_{ab} \textbf{H}^a \wedge \ba{b} &\!=\!& {\bf 0} \, ;\\
   \epsilon_{abc} \textbf{H}^a \wedge \ba{b} \wedge \ba{c} &\!=\!& {\bf 0} \, .\\ \nonumber
\end{eqnarray}

We see that the `electric part' of Einstein's equations, (\ref{LESoneform}, \ref{LEStwoform}), or (\ref{E1form}, \ref{E2form}), respectively, immediately reduce to those of Newtonian gravitation (up to the sign convention) for exact differential forms as geometrical limits of the coframes, i.e.,
\begin{equation}
   \ba{a} = \a{a}{k} \textbf{d}X^k \;\rightarrow \; \t{f}{a}{\mid k} \textbf{d}X^k \, .
\end{equation}
Applying this geometrical limit to the magnetic one--form (\ref{Hform}), we immediately see that it vanishes in the Newtonian limit, since the coefficients of the expansion one--forms $\boldsymbol{\Theta}^a = \dot{\boldsymbol{\eta}}^a$ reduce to the velocity gradient $\mbox{}^N \! \boldsymbol{\Theta}^a \rightarrow \dot{f}^a_{\;\; \mid k} \textbf{d}X^k = \t{v}{a}{,b} {\bf d}x^b$ with partial derivatives,
\begin{equation}
   \mbox{}^N \textbf{H}^a \;\rightarrow \; \t{\delta}{-}{be} \t{\epsilon}{acd}{} \t{v}{e}{,d,c} {\bf d}x^b = {\bf 0} \, ,
\end{equation}
where the Newtonian limit implies the existence of global Eulerian coordinates $x^b$. 
(For a useful reference on Cartan's formalism see \cite{nakahara}.)

%------------------------------------------------------------------------%
%------------------------------------------------------------------------%
\renewcommand{\theequation}{B.\arabic{equation}}
\setcounter{equation}{0} 
\section*{APPENDIX B: Other formulations involving electric and magnetic parts of the Weyl tensor}

We here provide other formulations in terms of the Weyl tensor parts that are helpful especially in future work.

Note that throughout this appendix we adopt the invariant volume element $\t{\varepsilon}{ikl}{}$ rather than the antisymmetric tensor $\t{\epsilon}{ikl}{}$, see (\ref{invvol}).

An interesting identity for the magnetic part is a relation of its covariant curl to covariant spatial derivatives of the expansion tensor,
\begin{equation}
   \t{g}{-}{im} \t{\varepsilon}{mkl}{} \t{H}{-}{jk \parallel l} = \t{\Theta}{{ \;\; \parallel k}}{\!\!\! ij \;\; \parallel k} - \t{\Theta}{k}{j \parallel i \parallel k} \: ,
\end{equation}
which together with (\ref{riccitheta}) allows us to relate the projected magnetic part of the Weyl tensor back to the time--derivative of the 3--Ricci tensor,
\begin{equation}
   \tdot{R}{-}{ij} = - 2 \t{g}{-}{m(i} \t{\varepsilon}{mkl}{} \t{H}{-}{j)k \parallel l} + \t{\Theta}{{\;\; \parallel k}}{\!\!\!ij \;\; \parallel k} - \t{\Theta}{-}{\parallel i \parallel j} \: . \label{randh}
\end{equation}
To rewrite (\ref{ADMorigthreedot}) in terms of the Weyl tensor we now derive a full set of evolution equations for the electric and magnetic parts.

%----------------------------------------------------------------%
\subsection*{Maxwell--like equations for the projected parts of the Weyl tensor}

The electric and magnetic parts (\ref{weylparts}) of the Weyl tensor obey a set of evolution equations similar to the Maxwell equations, see for example \cite{wainwrightellis,ellis:cargese}, \cite{bertschingerhamilton,kofmanpogosyan,ellisdunsby,maartensetal}. 
The Weyl tensor plays the role of the electrodynamical field tensor. Starting from the 4--Bianchi identities in the case of irrotational dust, $\t{u}{\mu}{}=(1,0,0,0)$ and $\t{u}{\mu}{; \nu}=\t{\Theta}{\mu}{\nu}$ (where a semicolon denotes $4-$covariant derivative),
\begin{equation}
   \t{C}{\mu}{\nu\kappa\lambda ; \mu} = 8 \pi G \left[ \left( \varrho \t{u}{-}{\nu} \t{u}{-}{[\lambda} \right)\!\mbox{}_{; \kappa]} + \frac{1}{3} \varrho_{; [\kappa} \t{g}{-}{\nu\lambda]} \right] \, ,
\end{equation}
we replace the Weyl tensor by its parts (\ref{weylparts}),
\begin{eqnarray}
   \t{C}{-}{\mu\nu\kappa\lambda} &\!\!=\!\!& \left( \t{g}{-}{\mu\nu\alpha\beta} \t{g}{-}{\kappa\lambda\gamma\delta} - \t{\varepsilon}{-}{\mu\nu\alpha\beta} \t{\varepsilon}{-}{\kappa\lambda\gamma\delta} \right) \t{u}{\alpha}{} \t{u}{\gamma}{} \t{E}{{\beta\delta}}{} \\
   &\!\!+\!\!& \left( \t{\varepsilon}{-}{\mu\nu\alpha\beta} \t{g}{-}{\kappa\lambda\gamma\delta} + \t{g}{-}{\mu\nu\alpha\beta} \t{\varepsilon}{-}{\kappa\lambda\gamma\delta} \right) \t{u}{\alpha}{} \t{u}{\gamma}{} \t{H}{{\beta\delta}}{} \, , \nonumber
\end{eqnarray}
where $\t{g}{-}{\mu\nu\alpha\beta} \equiv \t{g}{-}{\mu\alpha} \t{g}{-}{\nu\beta} - \t{g}{-}{\mu\beta} \t{g}{-}{\nu\alpha}$ and the tensor of the invariant volume element $\t{\varepsilon}{-}{\mu\nu\kappa\lambda} = \sqrt{-\mbox{}^{(4)}g} \t{\epsilon}{-}{\mu\nu\kappa\lambda}$ is defined analogous to (\ref{invvol}). We transform to Lagrangian coordinates as before and find, with $\sqrt{-\mbox{}^{(4)}g} = \sqrt{g} = J$ and $\t{\varepsilon}{-}{\mu\nu\kappa\lambda} \t{u}{\mu}{} = \t{\varepsilon}{-}{\nu\kappa\lambda}$ for the spatial invariant volume element, the following equations for the time--derivatives of the projected parts of the Weyl tensor or their covariant curls, respectively:
\begin{eqnarray}
   \tdot{E}{i}{j} &\!\!+\!\!& 2 \Theta \t{E}{i}{j} - \t{\Theta}{k}{j} \t{E}{i}{k} - \t{\Theta}{k}{l} \t{E}{l}{k} \t{\delta}{i}{j} - \t{\varepsilon}{ikl}{} \t{H}{-}{jl \parallel k} \nonumber \\
   &\!\!=\!\!& - 4 \pi G \frac{\mathring{J}}{J} \mathring{\varrho} \Big( \t{\Theta}{i}{j} - \frac{1}{3} \Theta \t{\delta}{i}{j} \Big), \label{Edotorig} \\
   \tdot{H}{i}{j} &\!\!+\!\!& 2 \Theta \t{H}{i}{j} - \t{\Theta}{k}{j} \t{H}{i}{k} - \t{\Theta}{k}{l} \t{H}{l}{k} \t{\delta}{i}{j} + \t{\varepsilon}{ikl}{} \t{E}{-}{jl \parallel k} \nonumber \\
   &\!\!=\!\!& - \frac{4 \pi G}{3} \frac{\mathring{J}}{J} \mathring{\varrho}_{\mid k} \t{\varepsilon}{ikl}{} \t{g}{-}{jl} \, . \label{Hdotorig}
\end{eqnarray}
Recalling that taking the time--derivative does not commute with the raising and lowering of indices, i.e.,
\begin{equation}
   \tdot{E}{-}{ij} = (\t{g}{-}{ik} \t{E}{k}{j})^{\dot{}} = \t{g}{-}{ik} \tdot{E}{k}{j} + 2 \t{\Theta}{-}{ik} \t{E}{k}{j}\;,
\end{equation}
(analogous for the magnetic part), and splitting the expansion tensor into its kinematical parts (for vanishing vorticity),
\begin{equation}
   \t{\Theta}{i}{j} = \t{\sigma}{i}{j} + \frac{1}{3} \theta \t{\delta}{i}{j}\;,
\end{equation}
we get the well--known forms of (\ref{Edotorig}), (\ref{Hdotorig}) as given in the literature, e.g. \cite{bertschingerhamilton}, \cite{ellisdunsby}.
For the covariant divergences of the Weyl tensor parts we find:
\begin{eqnarray}
   \t{E}{k}{i \parallel k} - \t{g}{-}{ik} \t{\varepsilon}{kmn}{} \t{\Theta}{-}{ml} \t{H}{l}{n} &\!\!=\!\!& \frac{8 \pi G}{3} \frac{\mathring{J}}{J} \mathring{\varrho}_{\mid i} \, \; ,\label{divE} \\
   \t{H}{k}{i \parallel k} + \t{g}{-}{ik} \t{\varepsilon}{kmn}{} \t{\Theta}{-}{ml} \t{E}{l}{n} &\!\!=\!\!& 0 \;\;.\label{divH}
\end{eqnarray}
The antisymmetric parts of Equations (\ref{Edotorig}) and (\ref{Hdotorig}) are equivalent to Equations (\ref{divE}) and (\ref{divH}), so we just take their symmetric parts (with lowered indices) into account, i.e.,
\begin{eqnarray}
   \tdot{E}{-}{ij} &\!\!+\!\!& 2 \Theta \t{E}{-}{ij} - 3 \t{\Theta}{-}{k(i} \t{E}{k}{j)} - \t{\Theta}{k}{l} \t{E}{l}{k} \t{g}{-}{ij} - \t{g}{-}{m(i} \t{\varepsilon}{mkl}{} \t{H}{-}{j)l \parallel k} \nonumber \\
   &\!\!=\!\!& - 4 \pi G \frac{\mathring{J}}{J} \mathring{\varrho} \Big( \t{\Theta}{-}{ij} - \frac{1}{3} \Theta \t{g}{-}{ij} \Big), \label{Edot} \\
   \tdot{H}{-}{ij} &\!\!+\!\!& 2 \Theta \t{H}{-}{ij} - 3 \t{\Theta}{-}{k(i} \t{H}{k}{j)} - \t{\Theta}{k}{l} \t{H}{l}{k} \t{g}{-}{ij} + \t{g}{-}{m(i} \t{\varepsilon}{mkl}{} \t{E}{-}{j)l \parallel k} \nonumber \\
   &\!\!=\!\!& 0 \, . \label{Hdot}
\end{eqnarray}
Note that for the right--hand--side of (\ref{Hdot}) we applied
\begin{equation}
   - \frac{4 \pi G}{3} \frac{\mathring{J}}{J} \mathring{\varrho}_{\mid k} \t{\varepsilon}{mkl}{} \t{g}{-}{m(i} \t{g}{-}{j)l} = 0 \; ,
\end{equation}
which is due to the antisymmetry of the Levi--Civita--tensor density.
These equations are automatically satisfied, if we insert the expressions for the parts of the Weyl tensor we found in the previous subsection and apply the ADM equations. Equation (\ref{Edot}) is the equivalent to (\ref{ADMorigthreedot}) we looked for.

%------------------------------------------------------------------------%
%------------------------------------------------------------------------%
\renewcommand{\theequation}{C.\arabic{equation}}
\setcounter{equation}{0} 
\section*{APPENDIX C: EXAMPLE FOR THE RZA}

In this appendix we shall derive a general solution for the trace evolution equation (\ref{tracefirstorder2}) for a flat FLRW background universe. (This way we also clarify the way in which the trace equations (\ref{rayfirstorder}) and (\ref{tracefirstorder2}) are equivalent.)
The homogeneous Friedmann equation that we restrict to the Einstein--de Sitter background, determines the scale factor 
\begin{equation}
   a(t) = \Big( \frac{t}{t_0} \Big)^{\!\frac{2}{3}} . \label{FLRWa}
\end{equation}
After solving the equations we restrict the solution to the Zel'dovich approximation and give some of the quantities discussed above for this particular choice of model.

%------------------------------------------------------------------------%
\subsection*{General solution for the first--order trace part}

For the inhomogeneous equation (\ref{tracefirstorder2}) we make the ansatz
\begin{equation}
   \t{P}{a}{i}(X,t) = \sum^{}_{\alpha} q_{\alpha}(t) \, \mbox{}^{\alpha}\t{Q}{a}{i}(X) \, .
\end{equation}
The time function $q(t)$ then has to obey the inhomogeneous differential equation
\begin{equation}
   \ddot{q} + \frac{2}{t} \dot{q} = \frac{1}{a^2} \Big( \ddot{q}(t_0) + \frac{2}{t_0} \dot{q}(t_0) \Big) . \label{FLRWdiffeqn}
\end{equation}
Since (\ref{FLRWdiffeqn}) is linear and of second order, the general solution can be written as superposition of three solutions, so $\alpha=1,2,p$. (Here $p$ denotes the particular solution of the inhomogeneous equations, whereas $\alpha=1,2$ label the homogeneous solutions.)
With the ansatz 
\begin{equation}
   q_{1/2}(t) = \Big( \frac{t}{t_0} \Big)^{\!n_{1/2}} , \quad q_p(t) = \Big( \frac{t}{t_0} \Big)^{\!p}
\end{equation}
we find
\begin{equation}
   n_1 = 0 \: , \quad n_2 = -1 \: , \quad p = \frac{2}{3} \: . \nonumber
\end{equation}
Hence,
\begin{equation}
   \t{P}{a}{i} = \, \mbox{}^{1}\t{Q}{a}{i} + \Big( \frac{t}{t_0} \Big)^{\!-1} \mbox{}^{2}\t{Q}{a}{i} + \Big( \frac{t}{t_0} \Big)^{\!\frac{2}{3}} \mbox{}^{p}\t{Q}{a}{i} \, . \label{FLRWperturbation}
\end{equation}
Using this we express the initial perturbation fields in terms of the initial conditions $\t{P}{a}{i}(t_0)$, $\tdot{P}{a}{i}(t_0)$ and $\tddot{P}{a}{i}(t_0)$,
\begin{eqnarray}
   \mbox{}^{1}\t{Q}{a}{i} &\!=\!& - \frac{3}{2} \tddot{P}{a}{i}(t_0) \, t_0^2 - 2 \, \tdot{P}{a}{i}(t_0) \, t_0 + \t{P}{a}{i}(t_0) \, , \nonumber \\
   \mbox{}^{2}\t{Q}{a}{i} &\!=\!& \;\;\: \frac{3}{5} \tddot{P}{a}{i}(t_0) \, t_0^2 + \frac{1}{5} \tdot{P}{a}{i}(t_0) \, t_0 \, , \nonumber \\
   \mbox{}^{p}\t{Q}{a}{i} &\!=\!& \; \frac{9}{10} \tddot{P}{a}{i}(t_0) \, t_0^2 + \frac{9}{5} \tdot{P}{a}{i}(t_0) \, t_0 \, . \nonumber
\end{eqnarray}
Thus, the general solution to (\ref{FLRWdiffeqn}) is, written in terms of the coframe for the deviation field:
\begin{eqnarray}
\label{FLRWdeformation}
   \mbox{}^{(1)}\t{\tilde{\eta}}{a}{i} &\!=\!& \t{\delta}{a}{i} + \t{P}{a}{i}(t_0) \\
   &\!+\!& 2 \left[ \frac{1}{10} \Big( \frac{t}{t_0} \Big)^{\!-1} + \frac{9}{10} \Big( \frac{t}{t_0} \Big)^{\!\frac{2}{3}} -1 \right] \! \tdot{P}{a}{i}(t_0) \, t_0 \nonumber \\
   &\!+\!& \frac{3}{2} \left[ \frac{2}{5} \Big( \frac{t}{t_0} \Big)^{\!-1} + \frac{3}{5} \Big( \frac{t}{t_0} \Big)^{\!\frac{2}{3}} -1 \right] \! \tddot{P}{a}{i}(t_0) \, t_0^2 \, . \nonumber
\end{eqnarray}

%------------------------------------------------------------------------%
\subsection*{The RZA for a flat FLRW background with $\Lambda = 0$}

As in section \ref{sect:rza} we write
\begin{equation*}
\t{P}{a}{i} \equiv \t{P}{a}{i}(X,t_0) \quad \textrm{and} \quad \tdot{P}{a}{i} \equiv \tdot{P}{a}{i}(X,t_0) \: ,
\end{equation*}
because $a(t)$, $\xi(t)$ are the only time--dependent functions.

First, we determine Zel'dovich's restriction in both forms (\ref{RZArestriction1}) and (\ref{RZArestriction2}) for the flat FLRW background with 
vanishing cosmological constant (Einstein--de Sitter background). 
The growing mode in (\ref{FLRWperturbation}) is $q(t) = (t/t_0)^{2/3}$, so with the scale factor (\ref{FLRWa}) we find
\begin{equation}
   \t{u}{a}{i}(t_0) = \t{w}{a}{i}(t_0) \, t_0 \, ,
\end{equation}
and, in terms of the initial perturbation fields,
\begin{equation}
   \tddot{P}{a}{i} \, t_0 = - \frac{1}{3} \tdot{P}{a}{i} \: .
\end{equation}
We subject the general first--order solution (\ref{FLRWdeformation}) to Zel'dovich's restriction and find, with
\begin{equation}
   \xi(t) = \frac{3}{2} \left[ \Big( \frac{t}{t_0} \Big)^{\!\frac{2}{3}} - 1 \right] t_0 \: ,
\end{equation}
for the ``peculiar''--coframe:
\begin{equation}
   \mbox{}^{\RZA}\t{\tilde{\eta}}{a}{i} = \t{\delta}{a}{i} + \t{P}{a}{i} + \frac{3}{2} \left[ \Big( \frac{t}{t_0} \Big)^{\!\frac{2}{3}} - 1 \right] \tdot{P}{a}{i} \, t_0 \: .
\label{FLRWcoframe}
\end{equation}
Apart from the term that arises because of the nonvanishing initial perturbation, this solution is familiar from the section on Newtonian dynamics above.
In the following we shall use $\xi$ and the $\mbox{}^{(n)\!\!}\t{J}{i}{j}$ of the RZA section to keep the equations short. 

Since $\t{\mathring{\eta}}{a}{i} = \t{\delta}{a}{i} + \t{P}{a}{i}$, we write
\begin{equation}
   \mbox{}^{\RZA} \a{a}{i} = \Big( \frac{t}{t_0} \Big)^{\!\frac{2}{3}} \big( \t{\mathring{\eta}}{a}{i} + \xi \tdot{P}{a}{i} \big) \, .
\end{equation}
Hence, the metric coefficients take the following form:
\begin{equation}
   \mbox{}^{\RZA} \t{g}{-}{ij} = \Big( \frac{t}{t_0} \Big)^{\!\frac{4}{3}} \t{\delta}{-}{ab} \big( \t{\mathring{\eta}}{a}{i} \t{\mathring{\eta}}{b}{j} + 2 \xi \t{\mathring{\eta}}{a}{(i} \tdot{P}{b}{j)} + \xi^2 \tdot{P}{a}{i} \tdot{P}{b}{j} \big) \, , \label{FLRWmetric}
\end{equation}
and the transformation determinant $\mbox{}^{\RZA\!} J$ reads:
\begin{equation}
   \mbox{}^{\RZA\!} J = \Big( \frac{t}{t_0} \Big)^{\!2} \, \mbox{}^{\RZA\!} \tilde{J}
    \: , \quad {\rm with} \quad \mbox{}^{\RZA\!}\tilde{J} \equiv \sum^{3}_{n=0} \xi^n \t{J}{-}{n} \: , 
\end{equation}
where $\t{J}{-}{n} \equiv \mbox{}^{(n)\!\!} \t{J}{k}{k}$ and $\tilde{J} \equiv \det (\t{\tilde{\eta}}{a}{i})$ is evaluated for the ``peculiar''--coframe field (\ref{FLRWcoframe}).

As in the general case, the nonlinearly evolved density can be calculated through its exact integral,
\begin{equation}
\label{FLRWrho}
   \mbox{}^{\RZA\!}\varrho = \frac{\mathring{\rho} \mathring{J}}{\mbox{}^{\RZA\!}J} = \frac{\varrho_H(t)}{\varrho_H(t_0)} \mathring{\varrho}(\vec{X}) \frac{\mathring{J}}{\mbox{}^{\RZA\!} \tilde{J}} \: .
\end{equation}

For the flat FLRW background the RZA expansion scalar and the mixed components of the RZA expansion tensor take the following forms:
\begin{equation}
   \mbox{}^{RZA} \Theta = \frac{2}{t} + \Big( \frac{t}{t_0} \Big)^{\!-\frac{1}{3}} \cdot \frac{\sum n \xi^{n-1} \t{J}{-}{n}}{\mbox{}^{\RZA\!} \tilde{J}} \;,
\end{equation}
and
\begin{equation}
\label{mixedTheta}
   \mbox{}^{RZA} \t{\Theta}{i}{j} =  \frac{2}{t} \cdot \frac{\sum \xi^{n} \mbox{}^{(n)\!\!}\t{J}{i}{j}}{\mbox{}^{\RZA\!} \tilde{J}} + \Big( \frac{t}{t_0} \Big)^{\!-\frac{1}{3}} \cdot \frac{\sum n \, \xi^{n-1} \mbox{}^{(n)\!\!}\t{J}{i}{j}}{\mbox{}^{\RZA\!} \tilde{J}} \,.
\end{equation}
(Here and in the following the summation is over $n=0\dots3\,$.)

The Riemann curvature tensor (\ref{RZARiemann}) then becomes:
\begin{equation}
   \mbox{}^{\RZA\!} \t{R}{i}{jkl} = 3 \cdot \frac{\sum \xi^n \big( \mbox{}^{(n)\!} \t{\tilde{R}}{i}{jkl} - \mbox{}^{(n)\!} \t{\tilde{R}}{i}{jlk} \big)}{\mbox{}^{\RZA\!} \tilde{J}}\;,
\end{equation}
and the Ricci tensor (\ref{RZARicci}) takes the form:
\begin{equation}
   \mbox{}^{\RZA\!} \t{R}{-}{ij} = 3 \cdot \frac{\sum \xi^n \big( \mbox{}^{(n)\!} \t{\tilde{R}}{k}{ikj} - \mbox{}^{(n)\!} \t{\tilde{R}}{k}{ijk} \big)}{\mbox{}^{\RZA\!} \tilde{J}} \, .
\end{equation}

Finally, we give the expressions for the parts of the Weyl tensor in the RZA. First, with
\begin{equation}
   H = \frac{2}{3t} \quad \Longrightarrow \quad \ddot{\xi} + 2H \dot{\xi} = \frac{1}{t_0} \Big( \frac{t}{t_0} \Big)^{-\frac{4}{3}}\;,
\end{equation}
the electric part reads:
\begin{eqnarray}
   \mbox{}^{RZA\!} \t{E}{i}{j} &\!\!=\!\!& \frac{2}{3t^2} \cdot \frac{\sum \xi^{n} \mbox{}^{(n)\!\!}\t{J}{i}{j}}{\mbox{}^{\RZA\!} \tilde{J}} - \frac{4 \pi G}{3} \cdot \frac{\varrho_H(t)}{\varrho_H(t_0)} \frac{\mathring{\rho} \mathring{J}}{\mbox{}^{\RZA\!} \tilde{J}} \, \t{\delta}{i}{j} \nonumber \\
   &\!\!-\!\!& \frac{1}{t_0} \Big( \frac{t}{t_0} \Big)^{\!-\frac{4}{3}} \cdot \frac{\sum n \, \xi^{n-1} \mbox{}^{(n)\!\!}\t{J}{i}{j}}{\mbox{}^{\RZA\!} \tilde{J}} \, , 
\end{eqnarray}
and the magnetic part:
\begin{equation}
   \mbox{}^{RZA\!} \t{H}{i}{j} = - \Big( \frac{t_0}{t} \Big) \cdot \frac{\t{\epsilon}{ikl}{} \left[ \tdot{P}{-}{kj \parallel l} + \t{\delta}{-}{ab} \big( \tdot{P}{a}{j} \t{P}{b}{k} + \xi \tdot{P}{a}{j} \tdot{P}{b}{k}\big)_{\parallel l} \right]}{\mbox{}^{\RZA\!} \tilde{J}} ,
\end{equation}
where $\parallel$ is the covariant derivative with respect to the RZA--metric (\ref{FLRWmetric}).

%------------------------------------------------------------------------%
%------------------------------------------------------------------------%

\end{document}